\documentclass[10pt,final,journal,a4paper,oneside,twocolumn]{IEEEtran}

\usepackage{subfigure}
\usepackage{srcltx}
\usepackage{psfrag}
\usepackage{epsfig}
\usepackage{color}
\usepackage[tbtags,sumlimits,nointlimits,reqno]{amsmath}
\usepackage{amssymb}
\usepackage{showkeys}   
\usepackage{color}
\usepackage{float}
\usepackage{mathtools, cuted}
\usepackage[makeroom]{cancel}
\begin{document}

\title{Finite-Difference Time-Domain (FDTD) Modelling of Space-Time Modulated Metasurfaces}

\author{%
       Scott A. Stewart, Tom. J. Smy and Shulabh~Gupta
\thanks{S. A.~Stewart, T. J. Smy and S.~Gupta, are with the Department of Electronics, Carleton University, Ottawa, Ontario, Canada. Email: shulabh.gupta@carleton.ca}
}
\markboth{MANUSCRIPT DRAFT}
{Shell \MakeLowercase{\textit{et al.}}: Bare Demo of IEEEtran.cls for Journals}

\maketitle
\begin{abstract}
A finite-difference time-domain (FDTD) modelling of finite-size zero thickness space-time modulated Huygens' metasurfaces based on Generalized Sheet Transition Conditions (GSTCs), is proposed and numerically demonstrated. A typical all-dielectric Huygens' unit cell is taken as an example and its material permittivity is modulated in both space and time, to emulate a travelling-type spatio-temporal perturbation on the metasurface. By mapping the permittivity variation onto the parameters of the equivalent Lorentzian electric and magnetic susceptibility densities, $\chi_\text{ee}$ and $\chi_\text{mm}$, the problem is formulated into a set of second-order differential equations in time with non-constant coefficients. The resulting field solutions are then conveniently solved using an explicit finite-difference technique and integrated with a Yee-cell based propagation region to visualize the scattered fields taking into account the various diffractive effects from the metasurface of finite size. Several examples are shown for both linear and space-time varying metasurfaces which are excited with normally incident plane and Gaussian beams, showing detailed scattering field solutions. While the time-modulated metasurface leads to the generation of new collinearly propagating temporal harmonics, these harmonics are angularly separated in space, when an additional space modulation is introduced in the metasurface. 
\end{abstract}

%



\section{Introduction}

Metasurfaces are the dimensional reduction of more general volumetric metamaterial structures, consisting of two dimensional arrays of sub-wavelength electromagnetic scatterers  \cite{Metasurface_Review}\cite{meta3}. By engineering the electromagnetic properties of the scattering particles, they can be used to control and engineer both the spatial wavefronts as well as the temporal shapes of the incident waves. They thus provide a powerful tool to transform incident fields into specified transmitted and reflected fields \cite{Metasurface_Synthesis_Caloz}\cite{BBParticlesTratyakov}\cite{Grbic_Metasurfaces}. They can either impart amplitude transformations, phase transformations or both, making them useful in diverse range of applications involving lensing, imaging \cite{meta2}\cite{meta3}, field transformations \cite{MetaFieldTransformation}, cloaking \cite{MetaCloak} and holograming \cite{MetaHolo}, for instance. Among several types of metasurfaces, \emph{Huygens' metasurfaces} have recently gathered a lot of attention due to their impedance matching capabilities with free-space. Huygen's metasurfaces are constructed using electrically small Huygen's sources exhibiting perfect cancellation of backscattering, due to optimal interactions of their electric and magnetic dipolar moments \cite{Kerker_Scattering}. Some efficient implementations of Huygens' metasurfaces are based on all-dielectric resonators \cite{Kivshar_Alldielectric}\cite{Elliptical_DMS}, and a good review can be found in \cite{AllDieelctricMTMS}.

Typical wave-shaping metasurfaces exhibit space-gradient designs, where the properties of the sub-wavelength scatterers are varied over the metasurface aperture. its simplest example is a generalized refracting metasurface \cite{GeneralizedRefraction}. Generalized refracting metasurfaces typically require periodic-space modulation of its constitutive parameters (susceptibility densities), and thus they may also be termed, alternatively, as \emph{space-modulated metasurfaces}. While most of the recent developments are in space-modulated metasurfaces, there has been a rapidly growing interest in space-time modulated metasurfaces, where the constitutive parameters of the metasurface are periodically varied in both space and time \cite{STGradMetasurface}\cite{ShaltoutSTMetasurface}. They are fundamentally different from reconfigurable metasurfaces, where the electromagnetic state of the metasurface can be tuned to achieve switching-type operations, such as in \cite{ReconfgMSoptics}. In contrast to re-configurable metasurfaces, the space-time modulated metasurfaces exhibit frequency conversions as a result of complex interactions of the incident wave with the metasurface at comparable time scales \cite{ShaltoutSTMetasurface}. In general, the principles of space-time modulated systems have been known for a long time \cite{OlinerST}\cite{TamirST} in the context of travelling-wave parametric amplifiers \cite{TWA} at microwaves and acousto-optic diffraction grating systems \cite{Goodman_Fourier_Optics} at optics \cite{Saleh_Teich_FP}\cite{TamirAcoustoDiffraction}, for instance. Combining the space-time modulation principles with the wave-front shaping capabilities of metasurfaces, represents an attractive enhancement of the metasurface functionalities. 

With this background and motivation, a space-time modulated Huygens' metasurface is rigorously modelled in the time-domain in this work, taking the exact space-discontinuity into account. This is achieved by applying the generalized sheet transition conditions (GSTCs) \cite{IdemenDiscont} in the time-domain to the space-time modulated Huygens' metasurface problem, and consequently solving for the output fields using the resulting explicit finite difference formulation. The general method has been recently proposed in \cite{Smy_Metasurface_Linear}, and is extended here to accommodate finite-size metasurfaces with space-time-varying susceptibilities of the metasurface. In this work, the surface susceptibilities of a Huygens' metasurface are extracted from a practical all-dielectric metasurface unit cell and then expressed in terms of a Lorentz dispersion for varying dielectric permittivities of the unit cell. By space-time-varying the parameters of the Lorentzian susceptibilities, an equivalent zero-thickness space-time modulated metasurface model is built. The space-time varying Lorentzian dispersions are next combined with GSTCs to form a complete set of matrix differential equations with non-constant coefficients. The resulting system of equations are then solved numerically using finite-difference time domain (FDTD) technique, where several cases of harmonic generation and spatial-spectral decomposition are demonstrated from a space-time modulated Huygens' metasurface.

The paper is organized as follows. Sec.~II describes the problem of field scattering from a space-time modulated Huygens' metasurface, in terms of electric and magnetic susceptibilities. Based on practical finite-thickness all-dielectric metasurface unit cells, various susceptibility functions are extracted and described in terms of Lorentzian forms for different material parameters. Sec.~III presents the governing field equations for an equivalent zero-thickness metasurface which are modelled using the extracted Lorentzian susceptibilities, and details the exact procedure and recipe to solve them numerically using finite-difference techniques. Sec.~IV presents numerical results for both non-modulated (linear-time-invariant) and space-time modulated Huygens' metasurface using the FDTD formulation of Sec.~II. Finally, the conclusions are provided in Sec. V. 

\section{Problem Formulation}

\subsection{Generalized Sheet Transition Conditions (GSTCs)}

A zero thickness metasurface, consisting of two dimensional arrays of sub-wavelength electromagnetic scatterers with zero thickness, is a space-discontinuity. The rigorous modelling of such discontinuities based on Generalized Sheet Transition Conditions (GSTCs) were developed by Idemen in \cite{IdemenDiscont}, which were later applied to metasurface problems in \cite{KuesterGSTC}. For a Huygens' metasurface lying in a $x-y$ plane, the modified Maxwell-Faraday and Maxwell-Ampere equations  can be written in the time-domain as,

\begin{subequations}\label{Eq:GSTC}
\begin{equation}
\hat{\mathbf{z}}\times\Delta \mathbf{H}(x,t) = \frac{d\mathbf{P}_{||}(x,t)}{dt}
\end{equation}
\begin{equation}
\Delta \mathbf{E}(x,t)\times \hat{\mathbf{z}} = \mu_0\frac{d\mathbf{M}_{||}(x,t)}{dt},
\end{equation}
\end{subequations}

\noindent where $\Delta \psi$ represents the differences between the fields on the two sides of the metasurface for all the vector component of the field $\psi$, i.e. $\mathbf{H}$ or $\mathbf{E}$ fields. The other terms $\mathbf{P}_{||}$ and $\mathbf{M}_{||}$ represent the electric and magnetic surface polarization densities, \emph{in the plane of the metasurface}, which depend on the total average fields around the metasurface. A more general description and discussion can be found in \cite{Metasurface_Synthesis_Caloz}, for interested readers.

If the input wave is assumed to be a $y-$polarized plane-wave, normally incident on the metasurface located at $z=0$, the differential fields (i.e. $\Delta \psi$) across the metasurface, can be written as

\begin{subequations}\label{Eq:FieldQuantities1}
\begin{equation}
\Delta\mathbf{E}(t) = \{E_t(t) - E_r(t) - E_0(t)\}~\hat{\mathbf{y}}
\end{equation}
\begin{equation}
\Delta\mathbf{H}(t) = -H_t(t) - H_r(t) + H_0(t)~\hat{\mathbf{x}},
\end{equation}
\end{subequations}

\noindent where $E_0(t)$ [$H_0(t)$] and $E_r(t)$ [$H_r(t)$] are the incident and reflected E-fields [H-fields] at $z=0_{-}$, and $E_t(t)$ [$H_t(t)$] is the transmitted field at $z=0_{+}$. The polarization densities induced on the metasurface in response to the incident fields, can be related to the average fields through surface susceptibilities, expressed in the frequency domain as

\begin{align}\label{Eq:FieldQuantities}
\mathbf{\tilde{P}}_{||}(\omega) &=  \epsilon_0\tilde{\chi}_\text{ee} \mathbf{\tilde{E}}_\text{av}(\omega), \quad
\mathbf{\tilde{M}}_{||}(\omega) &=  \tilde{\chi}_\text{mm} \mathbf{\tilde{H}}_\text{av}(\omega), 
\end{align}

\noindent where

\begin{align}
\mathbf{\tilde{E}}_\text{av} = \left[ \frac{\mathbf{\tilde{E}}_0 + \mathbf{\tilde{E}}_t + \mathbf{\tilde{E}}_r}{2}\right],\quad
\mathbf{\tilde{H}}_\text{av} = \left[ \frac{\mathbf{\tilde{H}}_0 + \mathbf{\tilde{H}}_t + \mathbf{\tilde{H}}_r}{2}\right],
\end{align}


\noindent and $\tilde{\chi}_\text{ee}(\omega)$ and $\tilde{\chi}_\text{mm}(\omega)$ are the electric and magnetic surface susceptibilities, respectively, which are assumed to be scalar. This describes a Huygen's source configuration consisting of orthogonal colocated electric and magnetic dipolar moments, modelled by $\tilde{\mathbf{P}}_{y}$ and $\tilde{\mathbf{M}}_{x}$, respectively. By properly controlling the electric and magnetic dipolar moments of a Huygens' surface, backscattering can be completely eliminated, resulting in a perfect transmission through the metasurface, which is naturally a useful functionality to have. A Huygens' metasurface thus can act as a perfect phase plate. For linear time-invariant metasurfaces, it can be shown that electric and magnetic surface susceptibilities are related to the total reflected and transmitted fields as \cite{Metasurface_Synthesis_Caloz}, 

\begin{subequations}\label{Eq:ChisFromS}
\begin{equation}
\tilde{\chi}_\text{ee}(\omega) =  \frac{2j}{k} \left(\frac{S_{21} + S_{11} - 1}{S_{21} + S_{11}+ 1} \right), 
\end{equation}
\begin{equation}
\tilde{\chi}_\text{mm}(\omega) =  \frac{2j}{k} \left(\frac{S_{21} - S_{11}- 1}{S_{21} - S_{11} + 1}\right),
\end{equation}
\end{subequations}

\noindent where $S_{21}(\omega) = \tilde{E}_t/\tilde{E}_i$ and $S_{11}(\omega) = \tilde{E}_r/\tilde{E}_i$. 

\begin{figure}[htbp]
\begin{center}
\psfrag{a}[c][c][0.8]{$z=0$}
\psfrag{z}[c][c][0.8]{$z$}
\psfrag{x}[c][c][0.8]{$x$}
\psfrag{b}[c][c][0.8]{$\psi(\mathbf{r}, t) = \psi_0(\mathbf{r}, t) \sin(\omega_0t)$}
\psfrag{f}[c][c][0.6]{$\boxed{\chi_\text{ee}(x, t),\; \chi_\text{mm}(x, t)}$}
\psfrag{d}[c][c][0.8]{$\psi_2(\mathbf{r}, t) \sin\{(\omega_0 + \omega_p)t\}$}
\psfrag{c}[l][c][0.8]{$\psi_1(\mathbf{r}, t) \sin\{\omega_0t\}$}
\psfrag{e}[c][c][0.8]{$\psi_3(\mathbf{r}, t) \sin\{(\omega_0 - \omega_p)t\}$}
\includegraphics[width=0.85\columnwidth]{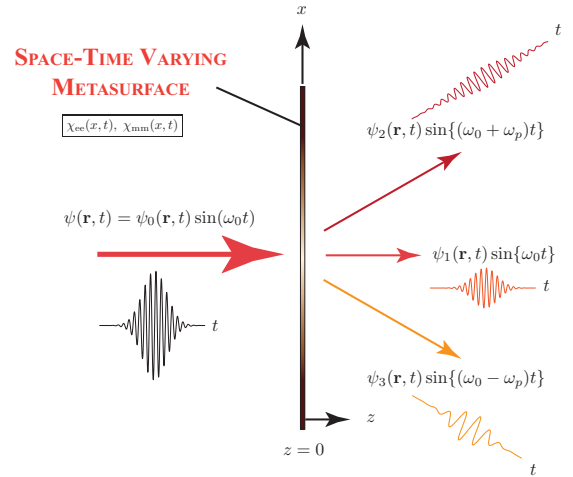}
\caption{A general Illustration of a space-time modulated Huygens' metasurface under normally incident pulsed plane-wave resulting in generation of several frequency harmonics, refracted along different angles. $\omega_p$ is the pumping frequency.}\label{Fig:Problem}
\end{center}
\end{figure}

In contrast to metasurfaces described using linear-time-invariant susceptibilities, Fig.~\ref{Fig:Problem} illustrates a time-variant Huygens' metasurface, whose susceptibilities are functions of space or time or both, i.e. $\chi_\text{ee}(x,t)$ and $\chi_\text{mm}(x,t)$. The space-time-variant metasurfaces, are naturally described in the time-domain.  When an arbitrary pulsed-wave $\psi(\mathbf{r}, t)$ is incident on the metasurface, its complex interaction with the metasurface generates new spectral components, which in turn are refracted along different directions, as illustrated in Fig.~\ref{Fig:Problem}. The objective of this work is to model such a zero-thickness space-time modulated Huygen's metasurface, for an arbitrary input signal, and determine the properties of the scattered fields in terms of new spatial (diffraction) and temporal frequency components. 

\begin{figure*}[htbp]
\begin{center}
\psfrag{a}[c][c][0.8]{frequency, $f$~(THz)}
\psfrag{b}[c][c][0.8]{S-parameters~(dB)}
\psfrag{c}[c][c][0.8]{electric susceptibility, $\tilde{\chi}_\text{ee}$}
\psfrag{d}[c][c][0.8]{magnetic susceptibility, $\tilde{\chi}_\text{mm}$}
\psfrag{e}[c][c][0.8]{$\Lambda_y$}
\psfrag{f}[c][c][0.8]{$\Lambda_x$}
\psfrag{g}[c][c][0.8]{$\epsilon_h$}
\psfrag{h}[c][c][0.8]{$\epsilon_r$}
\psfrag{i}[c][c][0.8]{unit cell \#1: $\tau = 0.9$}
\psfrag{j}[c][c][0.8]{unit cell \#2: $\tau = 0.7$}
\psfrag{k}[c][c][0.8]{unit cell \#3: $\tau = 0.5$}
\psfrag{m}[l][c][0.6]{Re\{$\cdot$\}}
\psfrag{n}[l][c][0.6]{Im\{$\cdot$\}}
\psfrag{p}[c][c][0.8]{$f_{0,1}$}
\psfrag{q}[c][c][0.8]{$f_{0,2}$}
\psfrag{r}[l][c][0.5]{$-3$~dB}
\psfrag{s}[l][c][0.5]{$-10$~dB}
\psfrag{t}[c][c][0.8]{$S_{11}$}
\psfrag{u}[c][c][0.8]{$d_1$}
\psfrag{v}[c][c][0.8]{$d_2$}
\psfrag{w}[c][c][0.8]{$\epsilon_r$}
\includegraphics[width=2\columnwidth]{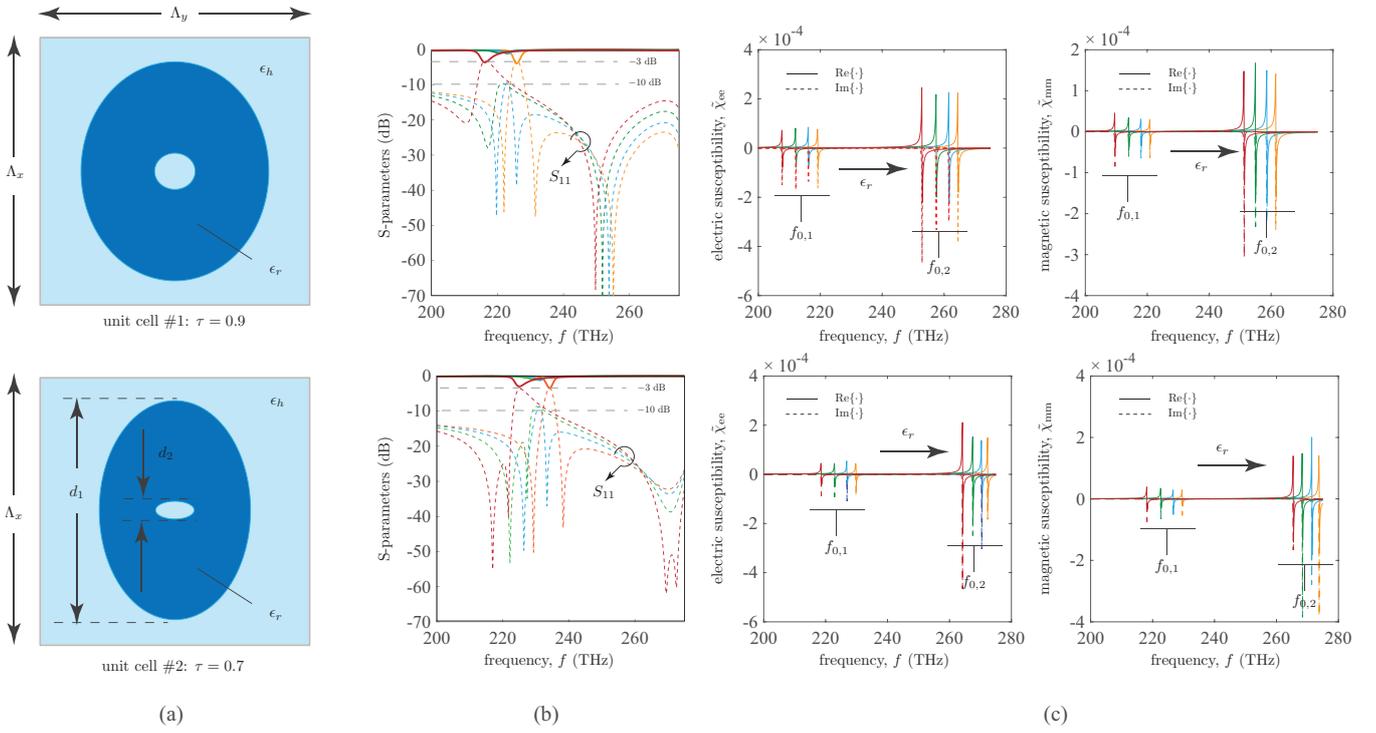}
\caption{All-dielectric unit cell for different material permittivities, $\epsilon_r = (11,\; 11.5,\; 12.2,\; 13)$. a) Two different unit cell geometries showing the transverse view with wave propagation along $z$. b) Transmission and reflection responses for varying permittivity values, and c) the corresponding extracted complex electric and magnetic susceptibilities, $\tilde{\chi}_\text{ee}$ and $\tilde{\chi}_\text{mm}$, respectively. The full-wave simulation parameters are: unit cell \#1 - $d_2 = 80$~nm, unit cell \#2 - $d_2 = 60$~nm. $d_1 = 560$~nm, substrate thickness $h=220$~nm and $\epsilon_h = 2.76$ for both cases. }\label{Fig:HFSSCell}
\end{center}
\end{figure*}

\subsection{Susceptibility Extraction}

Before a time-varying Huygens' surface can be modelled, its constitutive susceptibility functions must first be determined. A Huygen's metasurface is typically implemented using sub-wavelength all-dielectric unit cell sources. Fig.~\ref{Fig:HFSSCell}(a) shows one unit cell configuration, consisting of a holey dielectric resonator structure ($\epsilon_r$) embedded in a host medium ($\epsilon_h$). The unit cell size $\Lambda_x = \Lambda_y < \lambda_0$ to ensure the existence of a zeroth diffraction order only. The dielectric resonator is geometrically described in terms of its ellipticity ($\tau$), thickness ($h$) and the outer ($d_1$) and inner diameters ($d_2$) of the structure. By controlling the geometrical parameters, the Huygens' unit cell can be designed to achieve good transmission, with minimized reflection in a wide bandwidth. A convenient way to emulate a space-time modulated all-dielectric metasurface is by varying the permittivity, $\epsilon_r$, of the dielectric resonator, so that

\begin{equation}\label{Eq:EpsVary}
\epsilon_r(x,t) = \epsilon_r\{1 + \Delta_p \overbrace{\sin(\omega_pt - \beta_px)}^{f_m(t)}\},
\end{equation}

\noindent where $\omega_p$ is called the pumping frequency, $\beta_p$ is the spatial frequency of the perturbation in the permittivity and $\Delta_p$ is the modulation index. 

To determine the effect of permittivity on the transmission (reflection) response of the metasurface and estimate $\Delta_p$, a reference unit cell is first designed with a nominal value of $\epsilon_r = 11.9$ (silicon), exhibiting a quasi-perfect transmission through the metasurface in the entire frequency band of interest, i.e. a matched metasurface with $|S_{11}| < - 10$~dB. Any variation in $\epsilon_r$ is expected to increase the reflection from the metasurface compared to its optimum nominal value. Fig.~\ref{Fig:HFSSCell}(b) shows the FEM-HFSS simulated transmission and reflection responses of two different unit cell designs, for various dielectric permittivities around the nominal value of $\epsilon_r = 11.9$. It is found that when $11.5 < \epsilon_r < 12.2$, the metasurface still exhibits a good transmission with minimal reflections throughout the band. However, when this range is increased so that $11.50< \epsilon_r < 13.0$, a strong reflection is observed with a maximum value of $S_{11} = -3$~dB. This thus gives us two domains of $\Delta_p \approx  0.1$ and $\Delta_p \approx  0.05$, corresponding to a strong and weakly reflecting regimes.

Next, applying \eqref{Eq:ChisFromS} to the full-wave simulated transmission and reflection responses, the corresponding electric and magnetic susceptibilities are extracted for each unit cell and for different permittivity values, as shown in Fig.~\ref{Fig:HFSSCell}(c). To accurately quantify this behaviour, both $\tilde{\chi}_\text{ee}$ and $\tilde{\chi}_\text{mm}$, are approximated using a double-Lorentz function following 

\begin{subequations}\label{Eq:DualLZ}
\begin{equation}
	\tilde{\chi}_\text{ee}(\omega, \epsilon_r) = \sum_{i=1}^2\frac{\omega_{ep,i}^2}{(\omega_{e0,i}^2 - \omega^2) + i\alpha_{e,i}\omega}  
\end{equation}
\begin{equation}
	\tilde{\chi}_\text{mm}(\omega, \epsilon_r) = \sum_{i=1}^2\frac{\omega_{mp,i}^2}{(\omega_{m0,i}^2 - \omega^2) + i\alpha_{m,i}\omega}.
\end{equation}
\end{subequations}

\begin{figure}[htbp]
\begin{center}
\psfrag{b}[c][c][0.8]{resonant frequency, $f_0$~(THz)}
\psfrag{a}[c][c][0.8]{permittivity, $\epsilon_r$}
\psfrag{c}[c][c][0.8]{$f_{0,1}$}
\psfrag{A}[c][c][0.8]{$f_s$}
\psfrag{B}[c][c][0.8]{$\Delta f$}
\psfrag{d}[c][c][0.8]{$f_{0,2}$}
\psfrag{e}[l][c][0.8]{$f_\text{0,e}$}
\psfrag{f}[l][c][0.8]{$f_\text{0,m}$}
\psfrag{g}[c][c][0.8]{$|S_{11}| \approx -10$~dB}
\psfrag{h}[c][c][0.8]{$|S_{11}| \approx -3$~dB}
\includegraphics[width=\columnwidth]{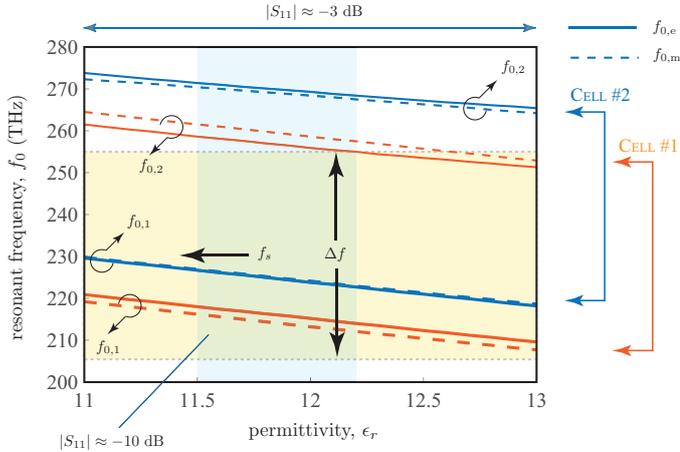}
\caption{The variation of the electric and magnetic resonant frequencies of the three metasurface unit cells of Fig.~\ref{Fig:HFSSCell}, as a function of material permittivity $\epsilon_r$, for both the first and second Lorentzian contributions. The center frequency and the bandwidth of the incident signal are $f_s = 230$~THz and $\Delta f \approx 50$~THz, respectively, to be used in numerical examples, for the rest of the paper, unless otherwise specified.}\label{Fig:W0vsEps}
\end{center}
\end{figure}

\noindent The extracted resonant frequencies of each unit cell, as a function of puck permittivity is shown in Fig.~\ref{Fig:W0vsEps}. It can be seen that both electric and magnetic susceptibilities require two resonant contributions. Furthermore, with decreasing permittivity, these resonances are shifted towards higher frequencies, as expected. A quasi-linear relationship between various resonant frequencies and permittivity $\epsilon_r$ is clearly seen which leads to an important conclusion: for a time-varying all-dielectric metasurface, any variation of permittivity can be equivalently modelled by varying the parameters (in particular, the resonant frequencies) of the Lorentzian susceptibilities using the same modulation function $f_m(t)$, i.e.

\begin{subequations}\label{Eq:W0VaryFunc}
\begin{equation}
\omega_{e0}(t) = \omega_{e0} + f_{e0}(t)  = \omega_{e0} + \Delta\omega_e \sin(\omega_pt ) 
\end{equation}
\begin{equation}
\omega_{m0}(t) = \omega_{e0} + f_{m0}(t)  = \omega_{m0} + \Delta\omega_m \sin(\omega_pt),
\end{equation}
\end{subequations}

\noindent where $\Delta\omega_{e,m}$ can be easily extracted from the slopes of the curves in Fig.~\ref{Fig:W0vsEps}. Finally, it should be noted that, in general, all the parameters of the Lorentzian susceptibilities (plasma frequencies $\omega_p$, resonant frequencies $\omega_0$ and loss coefficient, $\alpha$) depends on the permittivity $\epsilon_r$. However, for the units cells considered here, the variations in the plasma frequency and loss coefficient were found to be negligible, compared to that for resonant frequencies, and thus are ignored in the rest of the paper\footnote{While these parameters were found to be negligible, and thus kept equal to their normal static values to avoid cumbersome details here, they however can be easily taken into account in the proposed method at the expense of more complex modelling.}. Consequently, the problem of a computing the scattered fields from a time-varying (in general, space-time varying), finite-thickness all-dielectric metasurface for the given incident fields, can now be solved using a zero-thickness metasurface with Lorentzian electric and magnetic susceptibilities with space-time-varying resonant frequencies.

\section{Field Equations and Finite-Difference Formulation}

\subsection{Analytical Formulation of Space-Time Modulated Metasurfaces}

The relationship between the induced polarization densities and the exciting fields is described by \eqref{Eq:FieldQuantities} in the frequency domain, which for a monochromatic excitation, becomes a simple product in time-domain of $\chi$ and the corresponding temporal fields. However, for time-invariant metasurfaces, the problem is solved in the time-domain and thus the relationship between the polarization densities and the fields must be expressed as convolutions, i.e.

\begin{align}
\mathbf{P}_{||}(t) &= \epsilon_0 \chi_\text{ee}(t)\ast \left[\frac{\mathbf{E}_0(t) + \mathbf{E}_t(t) + \mathbf{E}_r(t)}{2}\right] \notag\\
\mathbf{M}_{||}(t) &=  \chi_\text{mm}(t)\ast  \left[ \frac{\mathbf{M}_0(t) + \mathbf{M}_t(t) + \mathbf{M}_r(t)}{2} \right].
\end{align}

Let us express the surface polarizabilities $\mathbf{P}_{||}$ and $\mathbf{M}_{||}$ in terms of average polarizations densities associated with incident, reflected and transmitted fields seperately, such that \footnote{For simplicity, all physical quantities are described here as function of time $t$ only for notational compactness. Wherever appropriate, they can be treated in general as functions of both space and time.}

\begin{subequations}
\begin{equation}
\mathbf{P}_{||}(t) = \frac{\mathbf{P}_0(t)+\mathbf{P}_t(t) +\mathbf{P}_r(t)}{2} 
\end{equation}
\begin{equation}
\mathbf{M}_{||}(t) = \frac{\mathbf{M}_0(t) +\mathbf{M}_t(t) +\mathbf{M}_r(t)}{2}
\end{equation}

\noindent where

\end{subequations}
\begin{align}
\mathbf{P}_i(t) = \mathcal{F}^{-1}\{{\mathbf{\tilde{P}}_i(\omega)}\} =  \mathcal{F}^{-1}\{\epsilon_0{\tilde{\chi}_\text{ee}(\omega)\mathbf{\tilde{E}}_i(\omega)}\}\notag\\
\mathbf{M}_i(t) = \mathcal{F}^{-1}\{{\mathbf{\tilde{M}}_i(\omega)}\} =  \mathcal{F}^{-1}\{{\tilde{\chi}_\text{mm}(\omega)\mathbf{\tilde{H}}_i(\omega)}\}
\end{align}

\noindent and the subscript $i=$0, t, and r for incident, transmitted and reflected fields, respectively. Since $\tilde{\chi}_\text{ee}$ and $\tilde{\chi}_\text{mm}$ exhibit Lorentz distributions following \eqref{Eq:DualLZ}, these relationships can be expressed in the time domain for time-varying resonant frequencies (and for each Lorentz contribution), as

\begin{subequations}\label{Eq:Set1}
\begin{equation}
\frac{d^2 Q_i}{dt^2} + \alpha_e \frac{dQ_i}{dt} + \omega_{e0}^2(t) Q_i = \omega_{ep}^2 E_i(t)
\end{equation}
\begin{equation}
 \frac{d^2 M_i}{dt^2} + \alpha_m\frac{dM_i}{dt} + \omega_{m0}^2(t) M_i =  \frac{\omega_{mp}^2}{\eta_0} E_i(t),\footnote{with RHS times (-1) for $E_0$ and $E_t$ and (+1) for $E_r$.}
\end{equation}
\end{subequations}

\noindent $Q_i = P_i/\epsilon_0$. Finally, from \eqref{Eq:GSTC}, under plane-wave excitations, 

\begin{subequations}\label{Eq:Set2}
\begin{equation}
  \frac{dM_t}{dt}   +  \frac{dM_r}{dt} + \frac{dM_0}{dt} = \frac{2}{\mu_0}(E_t - E_r - E_0)
\end{equation}
\begin{equation}
 \frac{dQ_t}{dt} +  \frac{dQ_r}{dt}  +  \frac{dQ_0}{dt} = \frac{2}{\eta_0\epsilon_0}(E_0-E_t - E_r).
\end{equation}
\end{subequations}

\noindent Equations~\eqref{Eq:Set1} and \eqref{Eq:Set2}, represent a total of eight field equations to be solved for two primary unknowns,  $E_t$, $E_r$, and six auxiliary unknowns, $P_t$, $P_r$, $P_0$, $M_t$, $M_r$, $M_0$, for a given input excitation field $E_0$. It should be noted that, while the metasurface susceptibilities exhibit double-Lorentz response, the formulation presented here is shown for only one Lorentzian contribution for simplicity. In reality, \eqref{Eq:Set1} represents 12 equations, 6 equations each for every Lorentzian term of $\tilde{\chi}$'s as described in \eqref{Eq:DualLZ}. Finally, the time-varying nature of the metasurface can be modelled using \eqref{Eq:W0VaryFunc}, for a given modulation function $f_m(t)$. This completes the mathematical formulation of the problem.

\subsection{Finite Difference Formulation}

The above set of equations can be conveniently solved by converting the second-order differential equations of  \eqref{Eq:Set1} into two first-order differential equations. For example, for electric polarization densities associated with incident fields, following \eqref{Eq:Set1}, we get \cite{Smy_Metasurface_Linear},\footnote{The time dependence, $(t)$, is dropped for compact notation.}

\begin{subequations}\label{Eq:AuxilliaryLZ}
\begin{equation}
\begin{split}
&\omega_{e0}\bar{Q}_0 = \frac{dQ_0}{dt} + \alpha_e Q_0, \\
& \frac{d\bar{Q}_0}{dt}  +  \bar{Q}_0\frac{1}{\omega_{e0}}\frac{d\omega_{e0}}{dt}  + \omega_{e0} Q_0= \frac{ \omega_{ep}^2}{\omega_{e0}} E_0
\end{split}
\end{equation}
\begin{equation}
\begin{split}
&\omega_{m0}\bar{M}_0 = \frac{dM_0}{dt} + \alpha_m M_0, \\
& \frac{d\bar{M}_0}{dt}  +  \bar{M}_0\frac{1}{\omega_{m0} }\frac{d\omega_{m0}}{dt}  + \omega_{m0} M_0 = -\frac{\omega_{mp}^2}{\eta_0\omega_{m0} } E_0.
\end{split}
\end{equation}
\end{subequations}

\noindent Similar set of equations can also be written for electric and magnetic polarizabilities, for both transmitted and reflected fields, as given in \eqref{Eq:MatrixP0}, \eqref{Eq:MatrixPt} and \eqref{Eq:MatrixPr}. These equations can now be written in a compact matrix form for incident, reflected and transmitted fields as 

\begin{figure*}[!t]
\begin{align}\label{Eq:MatrixP0}
	& \overbrace{\left[
	 \begin{array}{cccc}
		   1 & 0 & 0 & 0\\
		   0 & 1 & 0 &0 \\
		   0& 0& 1 & 0 \\
		  0 &0 &  0 & 1  \\
	\end{array} 
	\right]}^{\mathbf{W_1}^{4 \times 4}}
	\left[
	 \begin{array}{c}
		 Q_0'\\
		 \bar{Q}_0' \\
		 M_0'\\
		 \bar{M}_0'
	\end{array} 
	\right] +
	\overbrace{\left[
	 \begin{array}{cccc}
		\alpha_e & -\omega_{e0} & 0 & 0\\
		 \omega_{e0} &  \omega_{e0}'/\omega_{e0} & 0 & 0\\
		   0& 0& \alpha_m & -\omega_{m0}\\
		  0 & 0& \omega_{m0} & \omega_{m0}'/\omega_{m0}\\
	\end{array} 
	\right]}^{\mathbf{W_2(t)}^{4 \times 4}}
	\left[
	 \begin{array}{c}
		 Q_0\\
		 \bar{Q}_0 \\
		 M_0\\
		 \bar{M}_0
	\end{array} 
	\right]
	&=\overbrace{\left[
	 \begin{array}{c}
		 0\\
		  (\omega_{ep}^2/\omega_{e0})E_0\\
		 0\\
		 -(\omega_{mp}^2/\eta_0\omega_{m0})E_0
	\end{array} 
	\right]}^{\mathbf{E_1(t)}}
\end{align}

\begin{align}\label{Eq:MatrixPt}
	&\overbrace{\left[
	 \begin{array}{ccccc}
		   1 & 0 & 0 & 0 & 0\\
		   0 & 1 & 0 & 0 &0 \\
		   0 &0& 1& 0 & 0 \\
		  0 &0 &0 &  1 & 0 \\
	\end{array}
	\right]}^{\mathbf{T_1}^{4 \times 5}}
	\left[
	 \begin{array}{c}
		 Q_{t}'\\
		 \bar{Q}_{t}' \\
		 M_{t}'\\
		 \bar{M}_{t}'\\
		 E_{t}'
	\end{array} 
	\right] + 
	\overbrace{\left[
	 \begin{array}{ccccc}
		\alpha_e &  -\omega_{e0} & 0 & 0 & 0\\
		\omega_{e0} & \omega_{e0}'/\omega_{e0} &  0 & 0 & -\omega_{ep}^2/\omega_{e0}\\
		 0&  0& \alpha_m& -\omega_{m0} & 0\\
		   0 & 0 & \omega_{m0}& \omega_{m0}'/\omega_{m0} & (\omega_{mp}^2/\eta_0\omega_{m0})\\
	\end{array} 
	\right]}^{\mathbf{T_2(t)}^{4 \times 5}}
	\left[
	 \begin{array}{c}
		 Q_{t}\\
		 \bar{Q}_{t} \\
		 M_{t}\\
		 \bar{M}_{t}\\
		 E_{t}
	\end{array} 
	\right]
	&=\overbrace{\left[
	 \begin{array}{c}
		 0\\
		 0\\
		 0\\
		 0	
	\end{array} 
	\right]}^{\mathbf{E_2}}
\end{align}

\begin{align}\label{Eq:MatrixPr}
	&\overbrace{\left[
	 \begin{array}{ccccc}
		   1 & 0 & 0 & 0 & 0\\
		   0 & 1 & 0 & 0 &0 \\
		   0 &0& 1& 0 & 0 \\
		  0 &0 &0 &  1 & 0 \\
	\end{array}
	\right]}^{\mathbf{T_1}^{4 \times 5}}
	\left[
	 \begin{array}{c}
		 Q_{r}'\\
		 \bar{Q}_{r}' \\
		 M_{r}'\\
		 \bar{M}_{r}'\\
		 E_{r}'
	\end{array} 
	\right] + 
	\overbrace{\left[
	 \begin{array}{ccccc}
		\alpha_e &  -\omega_{e0} & 0 & 0 & 0\\
		\omega_{e0} & \omega_{e0}'/\omega_{e0} &  0 & 0 & -\omega_{ep}^2/\omega_{e0}\\
		 0&  0& \alpha_m& -\omega_{m0} & 0\\
		   0 & 0 & \omega_{m0}& \omega_{m0}'/\omega_{m0} & -(\omega_{mp}^2/\eta_0\omega_{m0})\\
	\end{array} 
	\right]}^{\mathbf{T_3(t)}^{4 \times 5}}
	\left[
	 \begin{array}{c}
		 Q_{r}\\
		 \bar{Q}_{r} \\
		 M_{r}\\
		 \bar{M}_{r}\\
		 E_{r}
	\end{array} 
	\right]
	&=\overbrace{\left[
	 \begin{array}{c}
		 0\\
		 0\\
		 0\\
		 0	
	\end{array} 
	\right]}^{\mathbf{E_2}}
\end{align}\rule[1ex]{18cm}{0.5pt}
\end{figure*}


\begin{subequations}\label{MatrixSet1}
\begin{equation}
	\mathbf{W_1}
	\left[
	 \begin{array}{c}
		 Q_0'\\
		 \bar{Q}_0' \\
		 M_0'\\
		 \bar{M}_0'
	\end{array} 
	\right] + \mathbf{W_2}(t)
	\overbrace{\left[
	 \begin{array}{c}
		 Q_0\\
		 \bar{Q}_0 \\
		 M_0\\
		 \bar{M}_0
	\end{array} 
	\right]}^{[\mathbf{V_0}]} = [\mathbf{E_1}(t)]\label{Eq:M0}
\end{equation}

\begin{equation}
	\mathbf{T_1}
	\left[
	 \begin{array}{c}
		 Q_{t}'\\
		 \bar{Q}_{t}' \\
		 M_{t}'\\
		 \bar{M}_{t}'\\
		 E_{t}'
	\end{array} 
	\right] + \mathbf{T_2}(t)
	\overbrace{\left[
	 \begin{array}{c}
		 Q_{t}\\
		 \bar{Q}_{t} \\
		 M_{t}\\
		 \bar{M}_{t}\\
		E_{t}
	\end{array} 
	\right]}^{[\mathbf{V_t}]} 
	= [\mathbf{0}],
\end{equation}

\begin{equation}
	\mathbf{T_1}
	\left[
	 \begin{array}{c}
		 Q_{r}'\\
		 \bar{Q}_{r}' \\
		 M_{r}'\\
		 \bar{M}_{r}'\\
		 E_{r}'\\		 
	\end{array} 
	\right] + \mathbf{T_2}(t)
	\overbrace{\left[
	 \begin{array}{c}
		 Q_{r}\\
		 \bar{Q}_{r} \\
		 M_{r}\\
		 \bar{M}_{r}\\
		 E_{r}
	\end{array} 
	\right]}^{[\mathbf{V_r}]} 
	= [\mathbf{0}],
\end{equation}
\end{subequations}

\noindent where matrix $[\mathbf{E_1}]$ includes the known excitation fields specified at the input of the metasurface. Similarly, the GSTC equations of \eqref{Eq:Set2} can be written in terms of new auxiliary variables as 

\begin{subequations}
\begin{equation}
\begin{split}
(\omega_{m0}\bar{M}_0 &- \alpha_m M_0)  + (\omega_{m0}\bar{M}_r - \alpha_m M_r )   \\
 & +  (\omega_{m0}\bar{M}_t - \alpha_m M_t) = \frac{2}{\mu_0} (E_t - E_r- E_0), 
 \end{split}
 \end{equation}
\begin{equation}
\begin{split}
(\omega_{e0}\bar{Q}_0 - \alpha_eQ_0)  &+ (\omega_{e0}\bar{Q}_r - \alpha_e Q_r ) +  (\omega_{e0}\bar{Q}_t - \alpha_e Q_t) \\
 &=  \frac{2}{\epsilon_0\eta_0}  (E_0 - E_t - E_r),
 \end{split}
 \end{equation}
\end{subequations}

\noindent to be further written in a compact matrix form as

\begin{align}\label{Eq:GSTC2Eq}
	\left[
	 \begin{array}{ccc}
	\mathbf{A_1} & \mathbf{A_2} &  \mathbf{A_3} \\
		\mathbf{B_1} & \mathbf{B_2} &  \mathbf{B_3}
	\end{array} 
	\right]_{2 \times 14  } [\mathbf{V}]_{14 \times 1}
		= \overbrace{\left[
	 \begin{array}{c}
	 -(2/\mu_0)E_0\\
	  (2/\epsilon_0\eta_0)E_0\\
	\end{array} 
	\right]}^{\mathbf{E_3}},
\end{align}

\noindent where $\mathbf{V}$ is a vector consisting of all the primary and auxiliary unknown variables, given by  $[ \mathbf{V}] = [ V_0, V_t, V_r]^\mathsf{T}$,  with $[\cdot]^\mathsf{T}$ as the matrix transpose and

\begin{align}
[\mathbf{A}_1] &= [0,\; 0,\; -\alpha_m,\; \omega_{m0}] \notag\\
[\mathbf{A}_2] &= [0,\; 0,\; -\alpha_m,\; \omega_{m0},\; -2/\mu_0] \notag\\
[\mathbf{A}_3] &= [0,\; 0,\; -\alpha_m,\; \omega_{m0},\; 2/\mu_0] \notag\\
[\mathbf{B}_1] &= [-\alpha_e,\; \omega_{e0},\;0,\; 0] \notag\\
[\mathbf{B}_2] &= [\mathbf{B}_3] = [-\alpha_e,\; \omega_{e0},\;0,\; 0,\;2/\epsilon_0\eta_0] \notag
\end{align}

\noindent Finally, using \eqref{MatrixSet1} and \eqref{Eq:GSTC2Eq}, we get 

\begin{align}\label{Eq:MatrixEquation}
	&\overbrace{\left[
	 \begin{array}{ccc}
	 	\mathbf{0} & \mathbf{0} &  \mathbf{0} \\
		\mathbf{0} & \mathbf{0} &  \mathbf{0}\\
		\mathbf{W_1} & \mathbf{0} &  \mathbf{0} \\
		\mathbf{0} & \mathbf{T_1} &  \mathbf{0} \\
		 \mathbf{0}  & \mathbf{0} &  \mathbf{T_1} \\
	\end{array} 
	\right]}^{\mathbf{[C]}}
	\frac{d[\mathbf{V}]}{dt}\notag\\
	&+ 
	\overbrace{\left[
	 \begin{array}{ccc}
	 	\mathbf{A_1} & \mathbf{A_2} &  \mathbf{A_3} \\
		\mathbf{B_1} & \mathbf{B_2} &  \mathbf{B_3}\\
		\mathbf{W_2}(t) & \mathbf{0} &  \mathbf{0} \\
		\mathbf{0} & \mathbf{T_2}(t) &  \mathbf{0} \\
		 \mathbf{0}  & \mathbf{0} &  \mathbf{T_2}(t) 
	\end{array} 
	\right]}^{\mathbf{[G(t)]}}
	[\mathbf{V}]=\mathbf{[E(t)]},
\end{align}

\noindent which can then further be written in compact notation as

\begin{equation}
\mathbf{[C]}\frac{d\mathbf{[V]}}{dt} + \mathbf{[G(t)]} \mathbf{[V]} = \mathbf{[E]},
\label{Eq:TVMetasurface_Equation}
\end{equation}

\noindent where $[ \mathbf{E}] = [ E_3, E_1, E_2, E_2]^\mathsf{T}$. In summary, for a specified time-domain field $E_0(x,t)$, included in $\mathbf{[E]}$ prescribed at the input of the metasurface at $z=0_-$, the above matrix equation provides the reflected fields $E_r(x,t)$ at $z=0_-$ and transmitted fields $E_t(x,t)$ at $z=0_+$, which are both included in the solution matrix $\mathbf{[V]}$. Matrices $\mathbf{[C]}$ and $\mathbf{[G]}$ contains the exact description of a given space-time modulated metasurface. For the trivial case of a static metasurface,  $\mathbf{[G(t)]} = \mathbf{[G]} $, so that \cite{Smy_Metasurface_Linear}

\begin{equation}
\mathbf{[C]}\frac{d\mathbf{[V]}}{dt} + \mathbf{[G]} \mathbf{[V]} = \mathbf{[E]}.
\label{Eq:Eq:StaticMetasurface_Equation}
\end{equation}

\noindent Finally, writing the explicit finite-difference form of \eqref{Eq:TVMetasurface_Equation} using the trapezoidal rule of integration, we get

\begin{align}\label{Eq:FDTDEquations}
\mathbf{[V]}_i    =  \left(\mathbf{[C]} +  \frac{\Delta t}{2}\mathbf{[G]}_i \right )^{-1} &\left[\Delta t\frac{ \mathbf{[E]_i} +  \mathbf{[E]_{i-1}}}{2} \right. \notag\\
& \left. + \left(\mathbf{[C]} - \frac{\Delta t}{2} \mathbf{[G]}_i\right) \mathbf{[V]}_{i-1}\right]
\end{align}

\noindent where $i$ is the index denoting the current time stamp.

\section{Numerical Demonstrations}

\subsection{Simulation Setup}

\begin{figure}[tbp]
\begin{center}
\psfrag{A}[c][c][0.8]{$E_r,\; H_r$}
\psfrag{B}[c][c][0.8]{$E_t,\; H_t$}
\psfrag{C}[c][c][0.8]{$E_0(x, z= 0_{-})$}
\psfrag{D}[c][c][0.8]{$E_t(x, z= 0_{+})$}
\psfrag{E}[c][c][0.8]{$z=0$}
\psfrag{F}[c][c][0.8]{$E_r(x, z= 0_{-})$}
\psfrag{X}[c][c][0.8]{$x$}
\psfrag{Z}[c][c][0.8]{$z$}
\psfrag{G}[c][c][0.6]{$\ell$}
\psfrag{a}[c][c][0.8]{$+z_0$}
\psfrag{b}[c][c][0.8]{$-z_0$}
\psfrag{c}[c][c][0.8]{$x_0$}
\includegraphics[width=0.85\columnwidth]{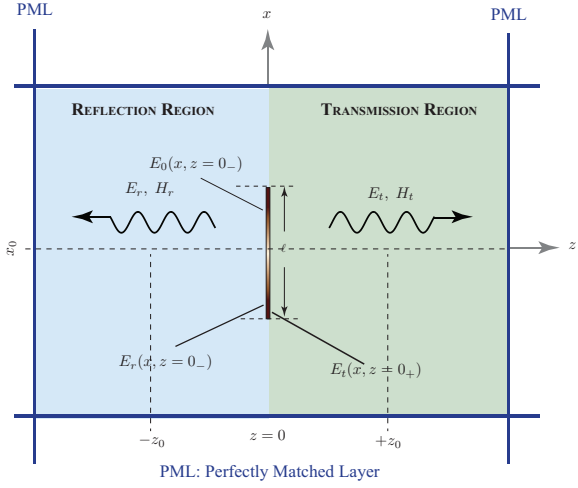}
\caption{Simulation setup showing the numerical domain including a finite-size metasurface of length $\ell$, and the surrounding boundary conditions. Shaded regions on the left and right represent the 2D discretized region where Maxwell's equations are solved on a conventional Yee cell.}\label{Fig:Setup}
\end{center}
\end{figure}

The developed FDTD formulation of a space-time modulated metasurface, determines the scattered fields just before and after a zero-thickness space-time varying metasurface, for a given time-domain input wave. To better visualize the scattered fields in space and time, and include the diffraction effects of a finite size metasurface, the FDTD equation of \eqref{Eq:FDTDEquations} can be solved on a conventional Yee-cell grid. Fig.~\ref{Fig:Setup} shows such a numerical setup consisting of a zero-thickness metasurface located at $z=0$ and of size $\ell$, which is considered much larger than the wavelength of the input signal. For simplicity, a 2D problem is considered here, so that any variations in the fields along the $y-$axis are assumed zero. The 2D computational domain is surrounded by Perfectly Matched Layers (PMLs) to eliminate any backscattering from the boundaries. A hard input source is specified for E-fields at $z=0_-$, which in general, is both a function of space and time, to model broadband time-domain pulses and arbitrary spatial wavefronts. The transmitted fields $E_t(x, z=0_+, t)$ generated from the metasurface, are first determined using \eqref{Eq:FDTDEquations} and subsequently forward-propagated in free-space on a conventional Yee-cell for $z > 0$ forming the transmission region. Similarly, the reflected fields  $E_r(x, z=0_-, t)$ are backward-propagated for $z<0$, forming the reflection region, as shown in Fig.~\ref{Fig:Setup}. The Yee-cell discritization is determined by the highest expected frequency and the time step set to satisfy the Courant-Friedrichs-Levy criteria \cite{taflove2000computational}.

The metasurface parameters are assumed to be the same as that in Fig.~\ref{Fig:HFSSCell} exhibiting the double-Lorentzian surface susceptibility functions, as obtained in Sec.~II-B. To emulate a space-time modulated metasurface, the resonant frequencies of each Lorentzian contributions are varied sinusoidally with a pumping frequency of $f_p$ and spatial frequency $\beta_p$. In all cases, the frequency of the incident signal in both the continuous-wave and pulsed regime (carrier frequency), is fixed to be $f_s = 230$~THz. The metasurface size is also fixed to $\ell = 25~\mu$m. The computation region is further limited to $\Delta x \times \Delta y =  50~\mu\text{m}\times 150~\mu\text{m}$, which was found sufficient to capture the scattered fields adequately. Finally, the Yee-cell region is discretized along $x-$ and $y-$directions using $n_x \times (n_z=3n_x)$ space samples, respectively\footnote{This ensures uniform meshing throughout the simulation domain. In general, a non-uniform meshing can be easily incorporated in the proposed method.}.

\begin{figure}[htbp]
\begin{center} 
\psfrag{a}[c][c][0.75]{$\epsilon_r(x,t) = \epsilon_r\{1 + \Delta_p \sin(\omega_pt - \beta_px)\}$}
\psfrag{b}[c][c][0.75]{$\omega_{e0}(\epsilon)$, $\omega_{m0}(\epsilon)$, $\omega_{ep}(\epsilon)$, $\omega_{mp}(\epsilon)$, $\alpha_e(\epsilon)$, $\alpha_m(\epsilon)~\forall~i$}
\psfrag{c}[c][c][0.75]{\shortstack{$\omega_{e0}(x,t)$, $\omega_{m0}(x,t)$, $\omega_{ep}(x,t)$, $\omega_{mp}(x,t)$, \\$\alpha_e(x,t)$, $\alpha_m(x,t)~\forall~i$}}
\psfrag{e}[c][c][0.75]{For each position $x = x'$ on the metasurface}
\psfrag{d}[c][c][0.75]{$\mathbf{[C]}$, $\mathbf{[G]}$}
\psfrag{f}[c][c][0.65]{$\mathbf{E}_0(x', t, z= 0_-)$}
\psfrag{h}[l][c][0.75]{$\mathbf{E}_t(x', t, z= 0_+)$}
\psfrag{j}[l][c][0.75]{$\mathbf{E}_r(x', t, z= 0_-)$}
\psfrag{m}[c][c][0.75]{For each time $t$}
\psfrag{k}[c][c][0.65]{$\displaystyle{\mathbf{[C]}\frac{d\mathbf{[V]}}{dt} + \mathbf{[G]} \mathbf{[V]} = \mathbf{[E]}}$}
\psfrag{n}[c][c][0.75]{$\mathbf{E}_{r,t}(x,z, t)$, $\mathbf{H}_{r,t}(x,z, t)$}
\includegraphics[width=0.9\columnwidth]{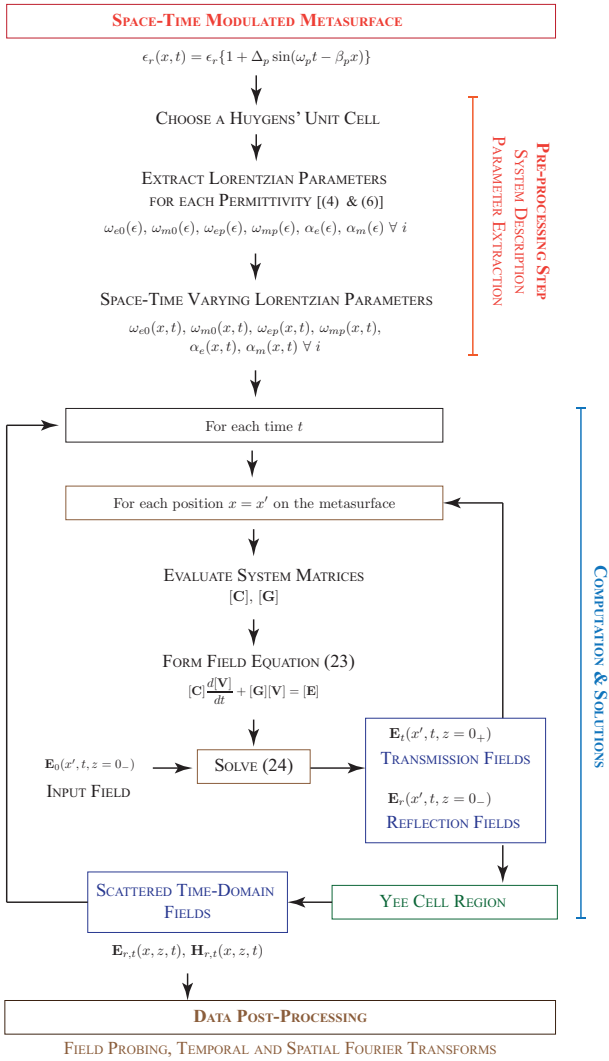}
\caption{Simulation flow chart illustrating the key-steps in solving a space-time modulated metasurface using the proposed method. }\label{Fig:FlowChart}
\end{center}
\end{figure}

Figure.~\ref{Fig:FlowChart} summarizes the key steps of the proposed method in solving a general space-time modulated metasurface. For a given Huygen's metasurface, the space-time modulation could be applied to the material permittivity, $\epsilon_r(x,t)$, as considered here. To model this surface, a unit cell is chosen (all-dielectric cell here) and full-wave simulated with varying permittivities within the range of $\epsilon_r \in [\text{min}\{\epsilon_r(x,t)\},\; \text{max}\{\epsilon_r(x,t)\}$. Using \eqref{Eq:ChisFromS}, their equivalent electric and magnetic susceptibilities, $\tilde{\chi}_\text{ee}(\epsilon_r)$ and $\tilde{\chi}_\text{mm}(\epsilon_r)$, are extracted and modelled using Lorentzian functions [\eqref{Eq:DualLZ}]. The variations in the various parameters of Lorentzian distributions are then recorded for each permittivity value (ex: Fig.~\ref{Fig:W0vsEps}). This mapping of the space-time modulated permittivity onto space-time modulated Lorentzian parameters complete the pre-processing step of the proposed method. Next, is the computation stage. At each time instants, various system matrices are evaluated, and the field equation \eqref{Eq:FDTDEquations} is solved for a given input field at each metasurface location. Once the entire reflected and transmitted fields are evaluated at $z = 0_-$ and $z = 0_+$, respectively, they are used as hard sources in the Yee-cell region to subsequently compute the scattered fields in the rest of computational domain. Finally, the scattered fields are post-processed, in both spatial and temporal frequency domains, to determine the harmonic generation and their angular refraction effects. 

\subsection{Linear-Time Invariant Metasurface}

\begin{figure*}[htbp]
\begin{center}
\psfrag{a}[c][c][0.6]{distance, $x$~($\mu$m)}
\psfrag{b}[c][c][0.6]{distance, $z$~($\mu$m)}
\psfrag{c}[c][c][0.6]{time, $t$~(fs)}
\psfrag{d}[c][c][0.6]{E-field, Re$\{E(x_0, z_0, t)\}$~V/m}
\psfrag{e}[c][c][0.6]{$k_x\times 10^{-6}$~(rad/m)}
\psfrag{f}[c][c][0.7]{$|\mathcal{F}_x\{E_t(x, +z_0, t)\}|^2$}
\psfrag{g}[c][c][0.7]{\color{white}{$E_y$}}
\psfrag{h}[c][c][0.7]{\color{white}{$H_x$}}
\psfrag{j}[c][c][0.7]{\color{white}{$H_y$}}
\psfrag{k}[c][c][0.6]{$20\log_{10}|\psi(x, z, t = 0.2~\text{ps})|$~dB}
\psfrag{q}[c][c][0.6]{$20\log_{10}|\psi(x, z, t = 0.6~\text{ps})|$~dB}
\psfrag{m}[c][c][0.6]{$E_0$}
\psfrag{n}[c][c][0.6]{$E_t$}
\psfrag{p}[c][c][0.6]{$E_r$}
\psfrag{r}[c][c][0.6]{Resonant frequency $\omega_{e0} \times 10^{-15}$~(rad/s)}
\psfrag{s}[c][c][0.6]{$k_{x0}$}
\psfrag{v}[c][c][0.7]{$\displaystyle{\theta_\text{theor.} = \sin^{-1}\left(\frac{\lambda_0}{\ell}\right)}$}
\psfrag{w}[c][c][0.6]{$\displaystyle{\theta_\text{sim}= \sin^{-1}\left(\frac{k_{x0}}{k_0}\right)}$}
\psfrag{x}[l][c][0.6]{simulation}
\psfrag{z}[c][c][0.6]{$k_x=0$}
\psfrag{y}[l][c][0.6]{theory}
\psfrag{F}[c][c][0.6]{Transmission Angle $\angle E_t(x)$~rad}
\psfrag{V}[c][c][0.7]{$\boxed{\chi_\text{ee} = \chi_\text{mm}}$}
\includegraphics[width=1.9\columnwidth]{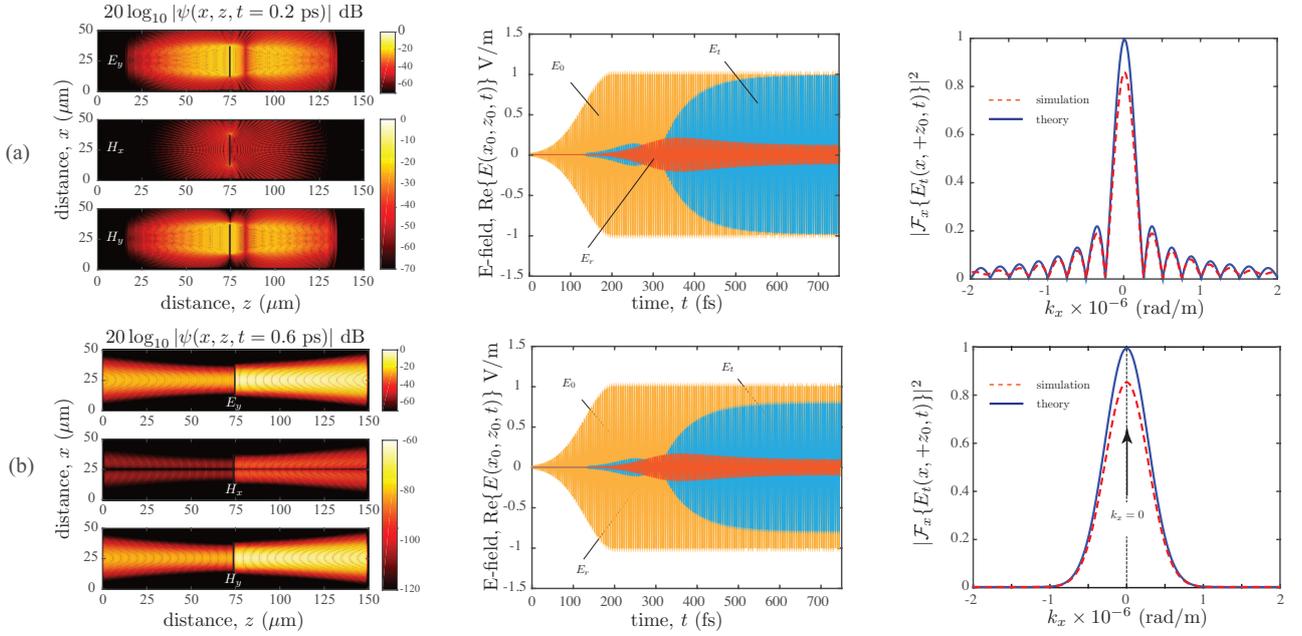}
\caption{Response of a linear time-invariant uniform metasurface, showing the instantaneous E-field and H-field distributions, corresponding time-domain reflected and transmitted fields at the field-probing locations $(x_0, \pm z_0)$, and the spatial Fourier transform of the fields in a transverse cross-section at $z= z_0$, for the two cases of (a) normally incident plane-wave, and (b) normally incident Gaussian-beam. The simulation parameters are as follows: CW frequency $f_s = 230$~THz, $\ell = 25~\mu$m, Gaussian field at the input $E_y(x, z=0_-) = \exp[-(x - x')^2/\sigma_x^2]$, with $x' = 25~\mu$m, $\sigma_x = 5~\mu$m. The rising Gaussian edge for plane-wave case is $\sigma_t = 100$~fs. Meshing $n_x =  400$. The metasurface parameters for are extracted from the unit cell~\#2 of Fig.~\ref{Fig:HFSSCell} : $f_{e0}= [224.63,\; 268.7950]$~THz, $f_{m0}= [224.4,\; 269.66]$~THz, $\omega_{ep}= [0.36,\; 0.95]$~Trad/s, $\omega_{em}= [0.29,\; 0.75]$~Trad/s, $\alpha_e = [500,\; 100]$~GHz and $\alpha_m = [100,\; 99]$~GHz. $\mathcal{F}_t\{\cdot\}$ and $\mathcal{F}_x\{\cdot\}$ denote temporal and spatial Fourier transforms, respectively.}\label{Fig:PW_GaussNonModMs}
\end{center}
\end{figure*}

Let us consider a linear time-invariant (LTI) (or non-modulated) metasurface first. The metasurface is assumed to be built using unit cell \#2 of Fig.~\ref{Fig:HFSSCell} with its equivalent Lorentzian parameters. Fig.~\ref{Fig:PW_GaussNonModMs}(a) shows the scattered fields when a continuous wave (CW) with a plane-wavefront is incident normally on the metasurface. Since unit cell \#2 exhibits a finite reflection (mismatched due to unequal $\chi_\text{ee}$ and $\chi_\text{mm}$), fields are manifest in both the transmission and reflection regions. Furthermore, since a plane-wave is assumed, the $x$ component of the H-field is negligible throughout, except at the two extremities of the metasurface capturing the fringing fields. The CW wave is switched on with a slowly rising Gaussian edge, and the corresponding time-domain fields at the centre of the reflection and transmission regions, at $(x_0, +z_0)$ and $(x_0, -z_0)$ respectively, are shown in the middle of Fig.~\ref{Fig:PW_GaussNonModMs}(a). The signal frequency $f_s = 230$~THz is close to the first Lorentzian resonant frequency, which leads to a strong interaction of the input wave with the metasurface, creating a distorted transition before, both reflected and transmitted signals reach a steady-state. To validate these results, the computed fields are compared with theory in the spectral ($k_x$) domain. This is achieved by computing the temporal Fourier transforms of $E_y(x, z= z_0, t)$ leading to $\tilde{E}_y(x, z= z_0, \omega)$, and then evaluating the spatial Fourier transform with respect to $x$ at the excitation frequency $\omega_0$, resulting in $\tilde{E}_y(k_x,z_0, \omega_0)$. This comparison is shown on the right of \ref{Fig:PW_GaussNonModMs}(a), where the theoretical distribution corresponds to the spatial Fourier transform of an ideal aperture (of size $\ell$) with constant field distribution. A good agreement is seen between the two. A smaller amplitude is seen for the computed spectrum, which can be attributed to dissipations losses of the metasurface and the energy being absorbed in the PMLs. The same procedure is repeated for a Gaussian beam and a step transition to CW as shown in \ref{Fig:PW_GaussNonModMs}(b), where a Gaussian beam diffracts through free-space after hitting the metasurface, again showing a good agreement with theoretical predictions.

\begin{figure}[htbp]
\begin{center} 
\psfrag{b}[c][c][0.6]{distance, $z$~($\mu$m)}
\psfrag{c}[c][c][0.6]{distance, $x$~($\mu$m)}
\psfrag{d}[c][c][0.6]{$20\log|E_y(x,z, t_1=0.2~\text{ps})|$~dB}
\psfrag{e}[c][c][0.6]{$20\log|E_y(x,z, t_2=0.4~\text{ps})|$~dB}
\psfrag{t}[c][c][0.6]{time, $t$~(fs)}
\psfrag{f}[c][c][0.6]{E-field, Re$\{E(x_0, \pm z_0, t)\}$~V/m}
\psfrag{h}[c][c][0.6]{Ave\{$|E(x,z, t)|^2$\}}
\psfrag{g}[l][c][0.4]{\shortstack{$n_x = 200$ \\ $n_x = 300$  \\ $n_x = 400$  \\ $n_x = 500$ \\ $n_x = 600$}}
\psfrag{m}[c][c][0.6]{$E_0(t)$}
\psfrag{n}[c][c][0.6]{$E_t(t)$}
\psfrag{p}[c][c][0.6]{$E_r(t)$}
\includegraphics[width=\columnwidth]{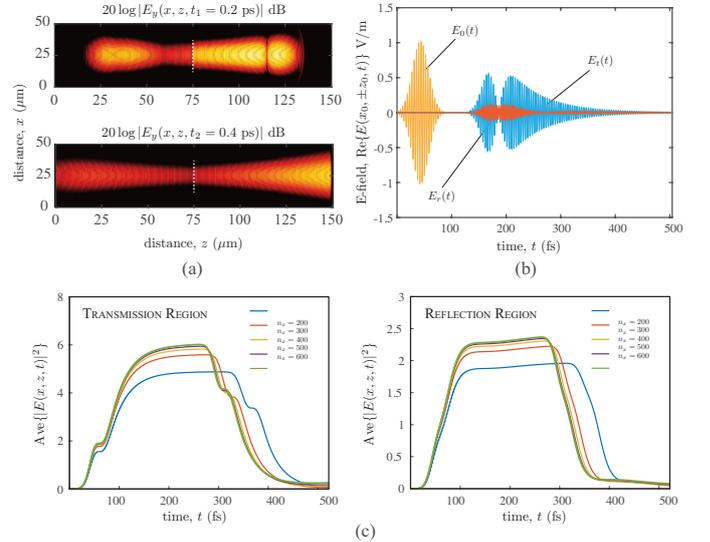}
\caption{Transmitted and reflection fields from a linear time-invariant metasurface for the case of a pulsed (broadband) Gaussian beam. (a) Spatial E-fields in the entire simulation region at two time instants $t_1=0.2$~ps, and $t_2 = 0.4$~ps. (b) Time-domain incident, reflected and transmitted fields at the probe locations in the transmission and reflection regions [$(x_0, -z_0)$ and $(x_0, +z_0)$], respectively. (c) The convergence plot of the simulation, where the average energy in the transmission and reflection regions converges at each time instant, for increasing value of the mesh density (related to $n_x$, where $n_z=3n_x$). The gaussian pulse width parameter is $\sigma_t = 50$~fs. }\label{Fig:GaussPulseNonModMS}
\end{center}
\end{figure}

Finally, one example of a pulsed (broadband Gaussian) Gaussian beam normally incident on a uniform metasurface of Fig.~\ref{Fig:PW_GaussNonModMs}, is shown in Fig.~\ref{Fig:GaussPulseNonModMS}. The Gaussian wavefronts are shown in Fig.~\ref{Fig:GaussPulseNonModMS}(a), at two time-instants showing the evolution of the pulsed wave, before it exits the simulation domain. The corresponding time domain fields are shown in Fig.~\ref{Fig:GaussPulseNonModMS}(b) for both reflected and transmitted fields. Due to the proximity of the carrier frequency of the Gaussian pulse to the resonant frequency of the metasurface, a strong temporal dispersion is seen with a slowly decaying pulse tail. Fig.~\ref{Fig:GaussPulseNonModMS}(c) shows the convergence of the simulation results, where the mesh grid in the Yee-cell region is continuously adapted by increasing $n_x$ until the total energy in the simulation region is converged to a stable value. It is found that $n_x > 500$ leads to a stable field solution, at each time instant. For all simulations throughout the paper, including that of Fig.~\ref{Fig:PW_GaussNonModMs}, the simulation convergence based on the total average energy in the simulation domain, is thoroughly checked and ensured.

\begin{figure*}[htbp]
\begin{center}
\psfrag{a}[c][c][0.8]{$x~\mu$m}
\psfrag{b}[c][c][0.8]{$z~\mu$m}
\psfrag{d}[c][c][0.8]{distance, $x$~($\mu$m)}
\psfrag{t}[c][c][0.6]{time, $t$~(fs)}
\psfrag{h}[c][c][0.6]{Ave\{$|E(x,z, t)|^2$\}}
\psfrag{c}[l][c][0.7]{\shortstack{$n =-2$ \\ $n = -1$  \\ $n = 0$  \\ $n = +1$ \\ $n = +2$}}
\psfrag{m}[c][c][0.6]{$E_0(t)$}
\psfrag{n}[c][c][0.6]{$E_t(t)$}
\psfrag{p}[c][c][0.6]{$E_r(t)$}
\psfrag{M}[c][c][0.8]{Power, $|\mathcal{F}_t\{E_r( x_0, z_0, t)\}|^2$~dB}
\psfrag{h}[c][c][0.8]{frequency, $f$~THz}
\psfrag{i}[c][c][0.8]{Power, $|\mathcal{F}_t\{E_t(x_0, z_0, t)\}|^2$~dB}
\psfrag{j}[c][c][0.8]{$f_s$}
\psfrag{e}[c][c][0.8]{E-fields~V/m}
\psfrag{K}[c][c][0.8]{$20\log_{10}|\psi(x, z, t = 0.8~\text{ps})|$~dB}
\psfrag{f}[c][c][0.8]{$|\mathcal{F}_t\{E_r(x, +z_0, t)|\}$}
\psfrag{F}[c][c][0.8]{$|\mathcal{F}_t\{E_t(x, +z_0, t)|\}$}
\psfrag{g}[c][c][0.8]{$f_p$}
\psfrag{G}[c][c][0.8]{\color{white}{$E_y$}}
\psfrag{H}[c][c][0.8]{\color{white}{$H_x$}}
\psfrag{J}[c][c][0.8]{\color{white}{$H_y$}}
\includegraphics[width=2\columnwidth]{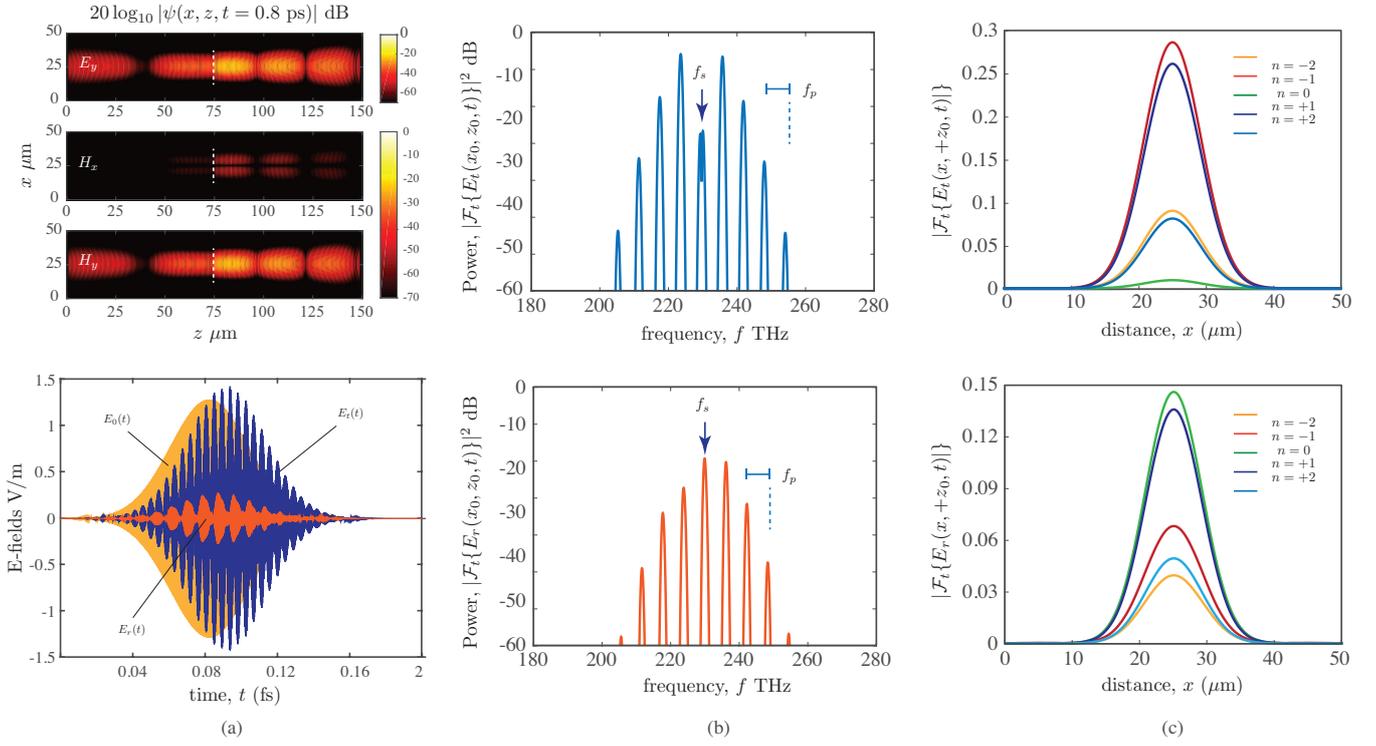}
\caption{Response of a time-only modulated (uniform) metasurface for a normally incident CW Gaussian beam with pumping frequency $f_m=0.025 f_s = 5.75$~THz and $\Delta_p = 0.1$, showing a) the time-domain E-fields at $(x_0, \pm z_0)$ and its spatial distribution in the transmission and reflection regions at a given time instant, (b) transmission and reflection temporal spectrum at $(x_0,\pm z_0)$, and (c) E-field distribution in the transverse plane at $z= \pm z_0$, corresponding to transmission and reflection fields, respectively. Only first few spectral components are shown, for clarity. $\mathcal{F}_t\{\cdot\}$ and $\mathcal{F}_x\{\cdot\}$ denote temporal and spatial Fourier transforms, respectively. The gaussian pulse width parameter is $\sigma_t = 200$~fs,}\label{Fig:TimeModulatedMS}
\end{center}
\end{figure*}

\subsection{Time-Modulated Metasurface}

Next, consider a time-modulated metasurface. Such a metasurface can be achieved by sinusoidally varying the permittivity of the all-dielectric unit cell, as described in Sec.~II-B, using \eqref{Eq:EpsVary} with $\beta_p = 0$, and $\Delta_p$ controlling the peak variation in the permittivity. For this example, consider again the unit cell\#2 of Fig.~\ref{Fig:HFSSCell} with its static permittivity of $\epsilon_r=11.9$ and $\Delta_p = 0.1$ to emulate the general case of strong reflection from the metasurface. Fig.~\ref{Fig:W0vsEps} shows the mapping of this permittivity variation on the resonant frequencies ($\omega_{e0}$ and $\omega_{m0}$) of the extracted Lorentzian susceptibilities, obtained using \eqref{Eq:DualLZ}, using which the time-modulated resonant frequencies of \eqref{Eq:W0VaryFunc} are determined. With this equivalent Lorentz description of this metasurface, the scattered fields can now be solved following the flow-chart of Fig.~\ref{Fig:FlowChart}.

Figure.~\ref{Fig:TimeModulatedMS} shows the scattered fields in both reflection and transmission, when excited with a  Gaussian pulsed beam, for an example pumping frequency of $\omega_p = 0.025\omega_s$ (2.5\% of the excitation frequency). Figure.~\ref{Fig:TimeModulatedMS}(a) shows the scattered fields in the transmission and reflection regions at a given time instant. The complete time-domain waveforms recorded at the centres of the transmission and reflection regions are also shown. The strong temporal beating indicates the generation of new frequency components, as expected from a time-modulated metasurface. Its temporal Fourier transform in Fig.~\ref{Fig:TimeModulatedMS}(b), shows the generation of equally spaced new spectral components at $\omega_n = \omega_0 \pm n \omega_p$ with different harmonic strengths. A strong frequency conversion is observed where the energy from the fundamental frequency $f_s$ is transferred equally to both up- and down-converted components in the transmission region. Interestingly, the strengths of these new spectral components are different in reflection as compared to that in transmission. This is however, also expected, since the transmission and reflection responses of a metasurface are different, in general for each spectral component. Further, the spatial profiles of first few of these harmonics at $z = \pm z_0$, are shown in Fig.~\ref{Fig:TimeModulatedMS}(c). All the newly generated harmonics are aligned, confirming that all the harmonics propagate together along the same direction ($\theta=0^\circ$), as expected from a uniform metasurface, i.e. a collinear propagation. 

 \begin{figure*}[htbp]
\begin{center}
\psfrag{a}[c][c][0.8]{distance, $x$~($\mu$m)}
\psfrag{b}[c][c][0.8]{distance, $z$~($\mu$m)}
\psfrag{d}[c][c][0.8]{Power, $|\mathcal{F}_x\{E_t(x, 0_+, \omega_n)\}|^2$}
\psfrag{e}[c][c][0.8]{Power, $|\mathcal{F}_x\{E_r(x, 0_-, \omega_n)\}|^2$}
\psfrag{k}[c][c][0.8]{$k_x\times 10^{-6}$~(rad/m)}
\psfrag{c}[c][c][0.8]{$|\mathcal{F}_t\{E_t(x, +z_0, t)|\}$}
\psfrag{g}[c][c][0.8]{$|\mathcal{F}_t\{E_r(x, +z_0, t)|\}$}
\psfrag{m}[c][c][0.8]{Harmonic index $n$}
\psfrag{f}[c][c][0.8]{Reflection angle $\theta_n$~deg}
\psfrag{j}[c][c][0.8]{Transmission angle $\theta_n$~deg}
\psfrag{n}[l][c][0.5]{\shortstack{$n =-2$ \\ $n = -1$  \\ $n = 0$  \\ $n = +1$ \\ $n = +2$}}
\psfrag{p}[l][c][0.5]{simulated}
\psfrag{q}[l][c][0.5]{Eq.~\eqref{Eq:BetapRefract}}
\psfrag{G}[c][c][0.8]{\color{white}{$E_y$}}
\psfrag{H}[c][c][0.8]{\color{white}{$H_x$}}
\psfrag{J}[c][c][0.8]{\color{white}{$H_y$}}
\psfrag{U}[c][c][0.8]{$20\log_{10}|\psi(x, z, t = 0.2~\text{ps})|$~dB}
\psfrag{S}[c][c][0.8]{$20\log_{10}|\psi(x, z, t = 0.6~\text{ps})|$~dB}
\psfrag{A}[c][c][0.5]{$n=-1$}
\psfrag{B}[c][c][0.5]{$n=+1$}
\psfrag{C}[c][c][0.5]{$n=0$}
\psfrag{E}[l][c][0.5]{$n=-2$}
\psfrag{D}[r][c][0.5]{$n=+2$}
\includegraphics[width=1.75\columnwidth]{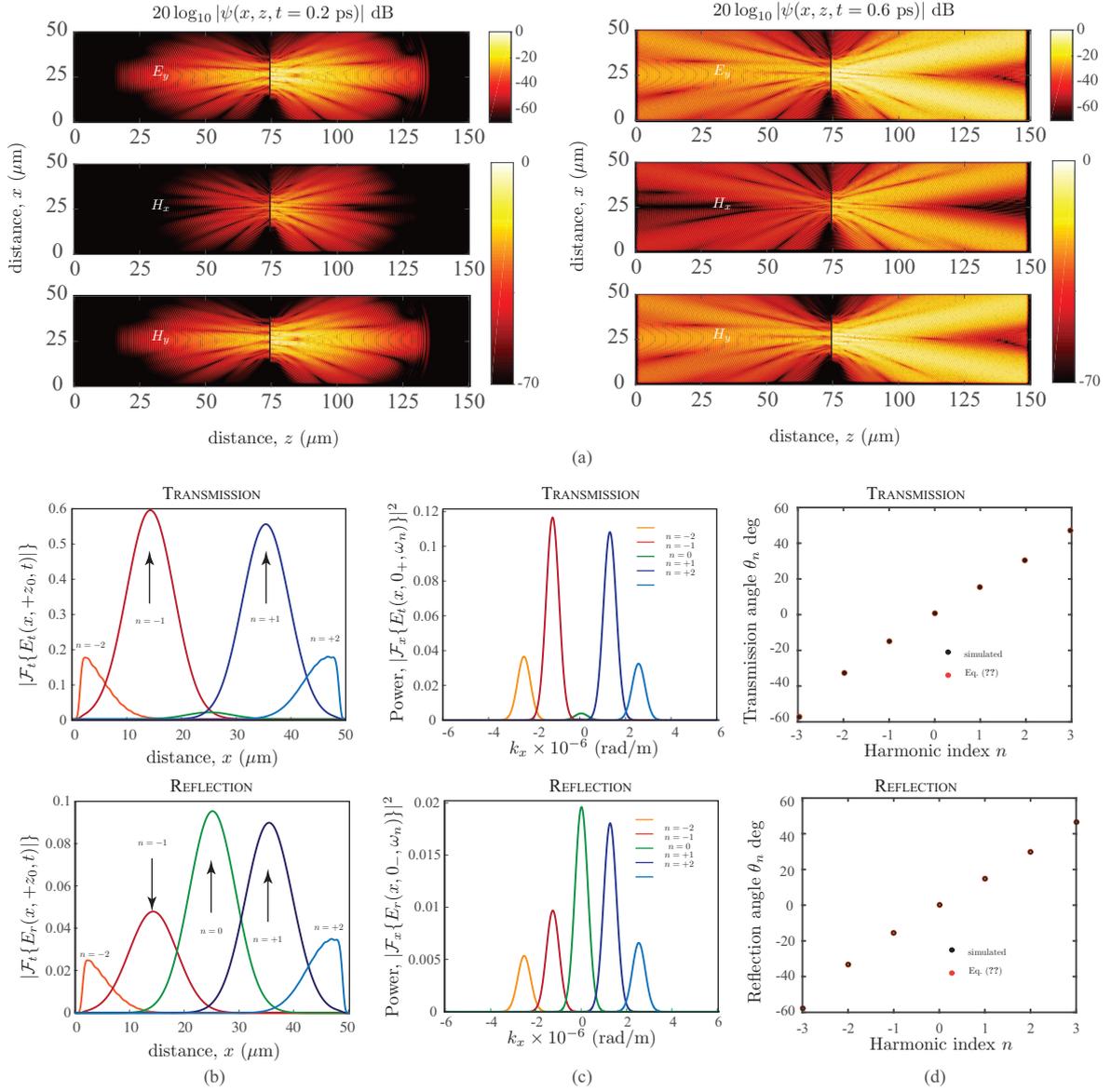}
\caption{Response of a space-time modulated metasurface with space-time varying permittivity according to \eqref{Eq:EpsVary}, showing (a) the scattered fields at two time instants. (b) Transmitted and reflected fields at $z=\pm z_0$ for individual harmonics, (c) the spatial Fourier transform of the transmitted and reflected fields at $z=0_{+}$ and $z=0_{-}$, respectively. (d) The refraction angles of each each harmonic components in reflection and transmission. The pumping frequency $f_p=0.025 f_s = 5.75$~THz and spatial modulation $\beta_p = 10\pi/\ell$. $\mathcal{F}_t\{\cdot\}$ and $\mathcal{F}_x\{\cdot\}$ denote temporal and spatial Fourier transforms, respectively.  }\label{Fig:SpaceTimeModulatedMS}
\end{center}
\end{figure*}

\subsection{Space-Time Modulated Metasurface}

Next consider a more general case of space-time modulated metasurfaces. Again, consider a uniform metasurface based on cell\#2 of Fig.~\ref{Fig:HFSSCell}, and vary the material permittivity, $\epsilon_r$ (and thus the resonant frequencies of the corresponding Lorentzian susceptibilities), in both space and time following \eqref{Eq:EpsVary}. Due to such periodic space-time perturbation of permittivity, the transmitted instantaneous fields just after the metasurface at $z=0_+$, can in general be expanded into a Floquet series as\footnote{This is valid strictly for an infinite metasurface, but provides a good estimation of the fields, as will be seen shortly. Similar expansion can be done for reflected fields with different set of coefficients $b_n$.}

\begin{equation}
E_t(x, z= 0_+, t) = \sum_{n=0}^\infty a_n e^{j\omega_n t} e^{j\beta_n x},
\end{equation}

\noindent where $\omega_n = \omega_s + n\omega_p$, and $\beta_n = \beta_{x0} + n\beta_p$ with $\beta_{x0}=0$ due to normal input incidence. Each harmonic term of this expansion with a temporal frequency $\omega_n$, represents an oblique forward propagating plane-wave along $+z$ direction, and making an angle $\theta_n$ measured from the normal of the metasurface, given by\footnote{Also valid for the temporal harmonics of the reflected fields.}

\begin{equation}\label{Eq:BetapRefract}
\theta(\omega_n) = \sin^{-1}\left\{\frac{n}{(1 + n\omega_p/\omega_s) }\frac{\beta_p}{k_0}\right\},
\end{equation}

\noindent where $k_0$ is the free-space wavenumber of the fundamental frequency. Similarly, the reflected fields can also be expanded in a Fourier series, with different set of harmonic amplitudes $b_n$, which will also be refracted along different angles. 

Figure.~\ref{Fig:SpaceTimeModulatedMS} shows a demonstration of a space-time modulated metasurface with a pumping frequency $f_p=0.025 f_s$ and spatial modulation $\beta_p = 10\pi/\ell$, for instance. Figure.~\ref{Fig:SpaceTimeModulatedMS}(a) shows the transmitted and reflected fields, at two time instants. Formation of several Gaussian beams travelling at different angles are clearly evident, in both reflection and transmission regions, as expected. Figure.~\ref{Fig:SpaceTimeModulatedMS}(b) shows the E-field distribution in the transmission and reflection region at $z=\pm z_0$, where various harmonics are seen to be clearly separated in space\footnote{The field distribution for $n = \pm 2$ harmonics are not fully seen as they are refracted into the PMLs and absorbed, due to large refraction angles.}. It should be noted that the fundamental frequency $f_s$ undergoes an almost complete down or up-conversion, consistent with the time-only modulated case of Fig.~\ref{Fig:TimeModulatedMS}. This suggests that the spatial frequency of the perturbation, $\beta_p$, in $\epsilon_r(x,t)$ has no effect on the strength of the newly generated harmonics, and it only controls their respective refraction angles following \eqref{Eq:BetapRefract}. Same observations can be made for the reflected fields. Fig.~\ref{Fig:SpaceTimeModulatedMS}(c) shows the spatial Fourier transforms of the transmitted and reflected fields just after and before the metasurface at $z=0_+$ and $z = z_-$, respectively, for the first few dominant harmonics. The locations $k_{x0, n}$ of the peaks of the $n^\text{th}$ harmonics are then used to compute the refraction angles using

\begin{equation}
\theta^\text{sim.}(\omega_n) = \sin^{-1}\left(\frac{k_{x0, n}}{k_n} \right),
\end{equation}

\noindent which are shown in Fig.~\ref{Fig:SpaceTimeModulatedMS}(d) for both reflection and transmission fields. An excellent agreement is observed with the theoretical predictions of \eqref{Eq:BetapRefract}.

\section{Conclusions}

An FDTD modelling of a finite-size zero thickness space-time modulated Huygens' metasurfaces has been proposed and numerically demonstrated using GSTCs. A typical all-dielectric Huygens' unit cell at optical frequencies, has been taken as an example and its material permittivity has been modulated in both space and time, to emulate a travelling-type spatio-temporal perturbation in the metasurface. By mapping the permittivity variation onto the parameters of the equivalent Lorentzian electric and magnetic susceptibilities, the problem is formulated into a set of second-order matrix differential equations with non-constant coefficients, which is then conveniently solved using an explicit finite-difference technique. The proposed techniques is then applied first to a linear-time-invariant metasurface with both plane-wave and Gaussian-beam excitations and then to both time-only and space-time modulated metasurface cases, presenting detailed scattering field solutions. While the time-modulated metasurface led to the generation of collinearly propagating temporal harmonics, these harmonics are further angularly separated in space, when a space modulation is introduced in the metasurface in addition to time. 

The proposed method is extremely versatile, flexible and capable of handling specified time-domain broadband signals with arbitrary spatial wavefronts. With physically motivated causal Lorentzian susceptibilities coupled through GSTCs, and user-defined space-time modulation functions, the proposed method represents a powerful and an efficient tool to accurately model and design sophisticated space-time modulated metasurfaces for complex electromagnetic wave control.

\section{Appendix}

\subsection{Extension to Dual-Lorentz Response}

Consider an electric surface susceptibility consisting of two Lorentzian contributions such that $\tilde{\chi}_\text{ee} = \tilde{\chi}_\text{ee,1} + \tilde{\chi}_\text{ee,2}$, following \eqref{Eq:DualLZ}. Its relates the total polarization density on the metasurface with the exciting E-fields, as $\tilde{P}_0(\omega) = \epsilon_0\tilde{\chi}_\text{ee}\tilde{E}_0(\omega)$. In this case, the total polarization can be broken into two parts as $\tilde{P}_0(\omega) = \tilde{P}_{01}(\omega) +  \tilde{P}_{02}(\omega)$, where $\tilde{P}_{01}(\omega) = \tilde{\chi}_\text{ee, 1}\tilde{E}_0(\omega)$ and $\tilde{P}_{02}(\omega) = \tilde{\chi}_\text{ee, 2}\tilde{E}_0(\omega)$, respectively. Similar decomposition can be done for the magnetic surface susceptibilities following an identical procedure. Now each electric and magnetic polarization term, can be written in the time domain using the Lorentzian form, for incident, transmitted and reflected fields, given in matrix form as

\small
\begin{align}
	\overbrace{\left[
	 \begin{array}{c c }
		 \mathbf{I} & \mathbf{0} \\
		 \mathbf{0} & \mathbf{I} \\
	\end{array}
	\right]}^{\mathbf{F_1}^{8\times 8}}
	[\mathbf{V_0}'] 
	+
	\overbrace{\left[
	 \begin{array}{c c }
		 \mathbf{L_1(1)} & \mathbf{0} \\
		 \mathbf{0} &  \mathbf{L_1(2)} \\
	\end{array}
	\right]}^{\mathbf{F_2}^{8\times 8}} [\mathbf{V_0}]  = 
	 \overbrace{\left[
	\begin{array}{c}
		 \mathbf{E_1(1)} \\
		 \mathbf{E_1(2)} \\
	\end{array}
	\right]}^{\mathbf{E_{Lz}}^{8\times 1}}
\end{align}

\small
\begin{align}
	\overbrace{\left[
	 \begin{array}{c c c}
		 \mathbf{I} & \mathbf{0} & \mathbf{0}\\
		 \mathbf{0} & \mathbf{I} & \mathbf{0}\\
	\end{array}
	\right]}^{\mathbf{F_3}^{8\times 9}}
	[\mathbf{V_t}'] 
	+
	\overbrace{\left[
	 \begin{array}{c c c }
		 \mathbf{L_1(1)} & \mathbf{0} & \mathbf{L_2^t(1)}\\
		 \mathbf{0} &  \mathbf{L_1(2)} & \mathbf{L_2^t(2)}\\
	\end{array}
	\right]}^{\mathbf{F_4}^{8\times 9}}[\mathbf{V_t}]  = [\mathbf{0}]^{8\times 1}
\end{align}

\small
\begin{align}
	\overbrace{\left[
	 \begin{array}{c c c}
		 \mathbf{I} & \mathbf{0} & \mathbf{0}\\
		 \mathbf{0} & \mathbf{I} & \mathbf{0}\\
	\end{array}
	\right]}^{\mathbf{F_5}^{8\times 9}}
	[\mathbf{V_r}'] 
	+
	\overbrace{\left[
	 \begin{array}{c c c }
		 \mathbf{L_1(1)} & \mathbf{0} & \mathbf{L_2^r(1)}\\
		 \mathbf{0} &  \mathbf{L_1(2)} & \mathbf{L_2^r(2)}\\
	\end{array}
	\right]}^{\mathbf{F_6}^{8\times 9}} [\mathbf{V_r}]  = [\mathbf{0}]^{8\times 1},
\end{align}

\normalsize
\noindent where the above matrices are given by

\begin{align}
\mathbf{L_1}(i)=
\left[
	 \begin{array}{cccc}
		\alpha_{ei} &  -\omega_{e0,i} & 0 & 0 \\
		\omega_{e0,i} & \frac{\omega_{e0,i}'}{\omega_{e0,i}} &  0 & 0 \\
		 0&  0& \alpha_{mi} & -\omega_{m0,i} \\
		   0 & 0 & \omega_{m0,i}& \frac{\omega_{m0,i}'}{\omega_{m0,i}} \\ 
	\end{array} 
	\right]
\end{align}

\begin{align}
\overbrace{\left[
	 \begin{array}{c}
		0\\
		 -\frac{\omega_{ep,i}^2}{\omega_{e0,i}}\\
		0\\
		   \frac{\omega_{mp,i}^2}{\eta_0\omega_{m0,i}}\\ 
	\end{array} 
\right]}^{\mathbf{L_2^t}(i)}, \quad
\overbrace{\left[
	 \begin{array}{c}
		0\\
		 -\frac{\omega_{ep,i}^2}{\omega_{e0,i}}\\
		0\\
		   -\frac{\omega_{mp,i}^2}{\eta_0\omega_{m0,i}}\\ 
	\end{array} 
	\right]}^{\mathbf{L_2^r}(i)},\quad
\overbrace{\left[
	 \begin{array}{c}
		0\\
		 \frac{\omega_{ep,i}^2}{\omega_{e0,i}}\\
		0\\
		 - \frac{\omega_{mp,i}^2}{\eta_0\omega_{m0,i}}\\ 
	\end{array} 
\right]}^{\mathbf{E_1}(i)}
\end{align}

\noindent and

\begin{align}
\mathbf{V_0} = 	\left[
	 \begin{array}{c}
		 Q_{01}\\
		 \bar{Q}_{01} \\
		 M_{01}\\
		 \bar{M}_{01}\\
		 Q_{02}\\
		 \bar{Q}_{02} \\
		 M_{02}\\
		 \bar{M}_{02}\\
	\end{array} 
	\right]
\mathbf{V_t} = 	\left[
	 \begin{array}{c}
		 Q_{t1}\\
		 \bar{Q}_{t1} \\
		 M_{t1}\\
		 \bar{M}_{t1}\\
		 Q_{t2}\\
		 \bar{Q}_{t2} \\
		 M_{t2}\\
		 \bar{M}_{t2}\\
		 E_{t}
	\end{array} 
	\right]
\mathbf{V_r} = 	\left[
	 \begin{array}{c}
		 Q_{r1}\\
		 \bar{Q}_{r1} \\
		 M_{r1}\\
		 \bar{M}_{r1}\\
		 Q_{r2}\\
		 \bar{Q}_{r2} \\
		 M_{r2}\\
		 \bar{M}_{r2}\\
		 E_{r}
	\end{array} 
	\right]
\end{align}

\noindent Finally, the two GSTC equations can also be written in terms of decomposed surface susceptibilities, as

\begin{subequations}
\begin{equation}
\begin{split}
(\omega_{m0,1}\bar{M}_{01} &- \alpha_{m1} M_{01})  + (\omega_{m0,1}\bar{M}_{r1} - \alpha_{m1} M_{r1} )   \\
 & +  (\omega_{m0,1}\bar{M}_{t1} - \alpha_{m1} M_{t1}) \\
 (\omega_{m0,2}\bar{M}_{02} &- \alpha_{m,2} M_{02})  + (\omega_{m0,2}\bar{M}_{r2} - \alpha_{m,2} M_{r2} )   \\
 & +  (\omega_{m0,2}\bar{M}_{t2} - \alpha_{m,2} M_{t2})
 = \frac{2}{\mu_0} (E_t - E_r- E_0), 
 \end{split}
 \end{equation}
\begin{equation}
\begin{split}
(\omega_{e0,1}\bar{Q}_{01} &- \alpha_{e1} Q_{01})  + (\omega_{e0,1}\bar{Q}_{r1} - \alpha_{e1} Q_{r1} )   \\
 & +  (\omega_{e0,1}\bar{Q}_{t1} - \alpha_{e1} Q_{t1}) \\
 (\omega_{e0,2}\bar{Q}_{02} &- \alpha_{e,2} Q_{02})  + (\omega_{e0,2}\bar{Q}_{r2} - \alpha_{e,2} Q_{r2} )   \\
 & +  (\omega_{e0,2}\bar{Q}_{t2} - \alpha_{e,2} Q_{t2})
 = \frac{2}{\epsilon_0\eta_0} (E_0 - E_t- E_r), 
 \end{split}
 \end{equation} 
 \end{subequations}

\noindent Combining the Lorentzian relations with the GSTC equations, lead to a matrix equation given by

\begin{align}\label{Eq:DualLZMatrix}
	[\mathbf{C}]
	&\left[
	 \begin{array}{c}
		 \mathbf{V_0}' \\
		 \mathbf{V_t}' \\
		 \mathbf{V_r}'
	\end{array}
	\right] +
	[\mathbf{G}]	%
	\left[
	 \begin{array}{c}
		 \mathbf{V_0} \\
		 \mathbf{V_t} \\
		 \mathbf{V_r}
	\end{array}
	\right]  = 
	\left[
	 \begin{array}{c}
		 \mathbf{E_{G}} \\
		  \mathbf{E_{Lz}} \\
		 \mathbf{0} \\
		  \mathbf{0}
	\end{array}
	\right]
\end{align}

\noindent where

\small
\begin{align}
	 [\mathbf{C}] = \left[\begin{array}{c c c}
	 	 \mathbf{0}   & \mathbf{0}  & \mathbf{0}\\
	 	 \mathbf{0}   & \mathbf{0}  & \mathbf{0}\\
		 \mathbf{F_1} & \mathbf{0} & \mathbf{0}\\
		 \mathbf{0} & \mathbf{F_3}  & \mathbf{0} \\
		 \mathbf{0} & \mathbf{0} & \mathbf{F_5} \\
	\end{array}\right],~
	 [\mathbf{G}] = \left[
	 \begin{array}{c c c}
		 \mathbf{J_1} & \mathbf{J_2} & \mathbf{J_3}\\
		 \mathbf{K_1} & \mathbf{K_2} & \mathbf{K_3}\\
		 \mathbf{F_2} & \mathbf{0} & \mathbf{0}\\
		 \mathbf{0} & \mathbf{F_4}  & \mathbf{0} \\
		 \mathbf{0} & \mathbf{0} & \mathbf{F_6} \\
	\end{array}
	\right]\notag
\end{align}

\begin{align}
&\mathbf{J_1} =
	\left[
	 \begin{array}{cc}
		 \mathbf{A_1}(1)  &   \mathbf{A_1}(2) 
	\end{array}
	\right], 	\mathbf{J_2} =
	\left[
	 \begin{array}{ccc}
		 \mathbf{A_1}(1)  &  \mathbf{A_1}(2) & -2/\mu_0
	\end{array}
	\right], \notag\\
&	\mathbf{J_3} =
	\left[
	 \begin{array}{ccc}
		 \mathbf{A_1}(1)  &  \mathbf{A_1}(2) & 2/\mu_0
	\end{array}
	\right],  \mathbf{A_1}(i) = [0, 0, -\alpha_{m,i}, \omega_{m,i}] \notag
\end{align}
\normalsize

\noindent and,

\small
\begin{align}
&\mathbf{K_1} =
	\left[
	 \begin{array}{cc}
		 \mathbf{B_1}(1)  &  \mathbf{B_1}(2) 
	\end{array}
	\right],
		\mathbf{K_2} =  
	\left[
	 \begin{array}{ccc}
		 \mathbf{B_1}(1)  &  \mathbf{B_1}(2) & 2/\epsilon_0\eta_0
	\end{array}
	\right], \notag\\
&	\mathbf{K_3} = \mathbf{K_2}, ~ 
\mathbf{B_1}(i) = [-\alpha_{e,i}, \omega_{e,i}, 0, 0],~
	\mathbf{E_G}  
	\left[
	 \begin{array}{c}
		 -(2/\mu_0)E_0\\
		 (2/\epsilon_0 \eta_0)E_0
	\end{array}
	\right].\notag
\end{align}
\normalsize

\noindent Equation~\ref{Eq:DualLZMatrix} can now be solved using standard numerical techniques. While this derivation is shown here for two Lorentzian contributions in the susceptibilities for illustration, it can easily be extended to arbitrary $N$ terms.

\bibliographystyle{IEEETran}
\bibliography{Stewart_ST_Metasurface_TAP_2016}

\end{document}